\documentclass[fleqn,usenatbib]{mnras}

\usepackage{newtxtext,newtxmath}

\usepackage[T1]{fontenc}

\DeclareRobustCommand{\VAN}[3]{#2}
\let\VANthebibliography\thebibliography
\def\thebibliography{\DeclareRobustCommand{\VAN}[3]{##3}\VANthebibliography}

\usepackage{graphicx}	
\usepackage{amsmath}	
\usepackage{multirow}
\usepackage{subfig}
\usepackage{multirow}
\usepackage{multirow}

\outer\def\gtae {$\buildrel {\lower3pt\hbox{$>$}} \over 
{\lower2pt\hbox{$\sim$}} $}
\outer\def\ltae {$\buildrel {\lower3pt\hbox{$<$}} \over 
{\lower2pt\hbox{$\sim$}} $}

\newcommand{\tess}{\it TESS}

\newcommand{\Porb}{P$_{\rm orb}$}

\newcommand{\chisq}{$\chi^{2}$}

\newcommand{\locccf}{CCF$_{\rm{loc}}$ }
\newcommand{\DIccf}{CCF$_{\rm{DI}}$ }
\newcommand{\locccfs}{CCFs$_{\rm{loc}}$ }
\newcommand{\DIccfs}{CCFs$_{\rm{DI}}$ }

\newcommand{\Nperiod}{\mbox{$5.44354215\substack{+0.00000307 \\ -0.00000297}$}}
\newcommand{\Ntc}{\mbox{$2458524.40869201\substack{+0.00030021 \\ -0.00029559}$}}  
\newcommand{\Nrprs}{\mbox{$0.05177\substack{+0.00063 \\ -0.00035}$}}
\newcommand{\Nars}{\mbox{$11.83\substack{+0.29 \\ -0.68}$}}
\newcommand{\Ninc}{\mbox{$88.85\substack{+0.74 \\ -0.94}$}}
\newcommand{\Nuonets}{\mbox{$0.3954\substack{+0.0023 \\ -0.0023}$}}
\newcommand{\Nutwots}{\mbox{$0.1519\substack{+0.0038 \\ -0.0038}$}}
\newcommand{\Nfzero}{\mbox{$-0.0000045\substack{+0.0000036 \\ -0.0000034}$}}
\newcommand{\Naau}{\mbox{$0.0668\substack{+0.0040 \\ -0.0044}$}}
\newcommand{\Nduration}{\mbox{$3.608\substack{+0.020 \\ -0.015}$}}  
\newcommand{\Nimpact}{\mbox{$0.213\substack{+0.169 \\ -0.153}$}}
\newcommand{\Nrpl}{\mbox{$0.6155\substack{+0.0306 \\ -0.0307}$}}  
\newcommand{\Nuonetsprior}{\mbox{$0.3955\pm0.0023$}}
\newcommand{\Nutwotsprior}{\mbox{$0.1523\pm0.0038$}}

\newcommand{\Ntcngone}{\mbox{$2459215.739534\pm0.0009405$}}
\newcommand{\Ntcngtwo}{\mbox{$2459264.729337\pm0.000633$}}
\newcommand{\Nuonengprior}{\mbox{$0.4531\pm0.0026$}}
\newcommand{\Nutwongprior}{\mbox{$0.1492\pm0.0042$}}

\title[Planetary Architecture and Stellar Variability of WASP-166]{The Hot Neptune WASP-166~b with ESPRESSO I: Refining the Planetary Architecture and Stellar Variability\thanks{Based on observations made at ESO's VLT (ESO Paranal Observatory, Chile) under ESO programme 106.21EM (PI Cegla) and utilising photometric lightcurves from the Transiting Exoplanet Survey Satellite (TESS) and the Next Generation Transit Survey (NGTS).}}

\author[L. Doyle et al.]{
L. Doyle,$^{1,2}$\thanks{E-mail: lauren.doyle@warwick.ac.uk}
H. M. Cegla,$^{1,2}$\thanks{UKRI Future Leaders Fellow}
E. Bryant,$^{1,2}$
D. Bayliss,$^{1,2}$
M. Lafarga,$^{1,2}$
D.~R.~Anderson$^{1,2}$
R. Allart,$^{3}$
\newauthor
V. Bourrier,$^{4}$
M. Brogi,$^{1,2,5}$
N.~ Buchschacher,$^{4}$ 
V. Kunovac,$^{6,7}$
M. Lendl,$^{4}$
C.~Lovis,$^{4}$
M.~Moyano,$^{8}$
\newauthor
N. Roguet-Kern,$^{4}$
J.~V.~Seidel,$^{9}$
D.~Sosnowska,$^{4}$
P.~J.~Wheatley,$^{1,2}$
J.~S. Acton,$^{10}$
M.~R. Burleigh,$^{10}$
\newauthor
S.~L. Casewell,$^{10}$
S. Gill$^{1,2}$
M.~R. Goad,$^{10}$ 
B.A.~Henderson,$^{10}$
J. S. Jenkins,$^{11,12}$
R.~H. Tilbrook,$^{10}$
\newauthor
and R.~G.~West$^{1,2}$
\\
$^{1}$Centre for Exoplanets and Habitability, University of Warwick, Coventry, CV4 7AL, UK \\
$^{2}$Department of Physics, University of Warwick, Coventry, CV4 7AL, UK\\
$^{3}$ Department of Physics, and Institute for Research on Exoplanets, Universit\'e de Montr\'eal, Montr\'eal, H3T 1J4, Canada \\
$^{4}$Observatoire Astronomique de l'Universit\'e de Gen\`eve, Chemin Pegasi 51b, CH-1290 Versoix, Switzerland \\
$^{5}$INAF – Osservatorio Astrofisico di Torino, Via Osservatorio 20, 10025, Pino Torinese, Italy \\
$^{6}$Lowell Observatory, 1400 W. Mars Hill Rd., Flagstaff, AZ 86001, USA \\
$^{7}$School of Physics and Astronomy, University of Birmingham, Edgbaston, Birmingham B15 2TT, UK \\
$^{8}$Instituto de Astronom\'ia, Universidad Cat\'olica del Norte,Angamos 0610, 1270709, Antofagasta, Chile\\
$^{9}$European Southern Observatory, Alonso de C\'ordova 3107, Vitacura, Regi\'on Metropolitana, Chile\\
$^{10}$ School of Physics and Astronomy, University of Leicester, Leicester LE1 7RH, UK\\
$^{11}$N\'ucleo de Astronom\'ia, Facultad de Ingenier\'ia y Ciencias, Universidad Diego Portales, Av. Ej\'ercito 441, Santiago, Chile\\
$^{12}$Centro de Astrof\'isica y Tecnolog\'ias Afines (CATA), Casilla 36-D, Santiago, Chile\\
}

\date{Accepted XXX. Received YYY; in original form ZZZ}

\pubyear{2022}

\begin{document}
\label{firstpage}
\pagerange{\pageref{firstpage}--\pageref{lastpage}}
\maketitle

\begin{abstract}
In this paper, we present high-resolution spectroscopic transit observations from ESPRESSO of the super-Neptune WASP-166~b. In addition to spectroscopic ESPRESSO data, we analyse photometric data from {\tess} of six WASP-166~b transits along with simultaneous NGTS observations of the ESPRESSO runs. These observations were used to fit for the planetary parameters as well as assessing the level of stellar activity (e.g. spot crossings, flares) present during the ESPRESSO observations. We utilise the Reloaded Rossiter McLaughlin (RRM) technique to spatially resolve the stellar surface, characterising the centre-to-limb convection-induced variations, and to refine the star-planet obliquity. We find WASP-166~b has a projected obliquity of $\lambda = -15.52^{+2.85}_{-2.76}$$^{\circ}$ and $v\sin(i) = 4.97 \pm 0.09$~kms$^{-1}$ which is consistent with the literature. We were able to characterise centre-to-limb convective variations as a result of granulation on the surface of the star on the order of a few kms$^{-1}$ for the first time. We modelled the centre-to-limb convective variations using a linear, quadratic and cubic model with the cubic being preferred. In addition, by modelling the differential rotation and centre-to-limb convective variations simultaneously we were able to retrieve a potential anti-solar differential rotational shear ($\alpha \sim$ -0.5) and stellar inclination ($i_*$ either 42.03$^{+9.13}_{-9.60}$$^{\circ}$ or 133.64$^{+8.42}_{-7.98}$$^{\circ}$ if the star is pointing towards or away from us). Finally, we investigate how the shape of the cross-correlation functions change as a function of limb angle and compare our results to magnetohydrodynamic simulations.  
\end{abstract}




\section{Introduction}
Stellar magnetic activity and surface phenomena, such as flares, spots, differential rotation (DR) and centre-to-limb convective velocity variations (CLV), can alter the observed stellar line profiles and induce Doppler shifts which can cause biases in the calculations of planetary companion properties \citep[e.g.][]{saar1997activity, oshagh2016can, cegla2016modeling, cegla2016rossiter}. Furthermore, DR plays a critical role in dynamo processes which are largely responsible for the generation of magnetic fields \citep[e.g.][]{kitchatinov2011differential, karak2020stellar}. Therefore, understanding these phenomena is not just important for exoplanet characterisation, but for magnetic activity as a whole. 

When a planet transits a host star, a portion of the starlight is blocked in the line of sight and a distortion of the velocities is observed, known as the Rossiter-McLaughlin (RM) effect (see \citet{rossiter1924detection, mclaughlin1924some} for original studies and \citet{queloz2000detection} for the first exoplanet case). The Reloaded RM (RRM) technique \citep[see][]{cegla2016rossiter} isolates the blocked starlight behind the planet to spatially resolve the stellar spectrum. The isolated starlight from the RRM can be used to derive the projected obliquity, $\lambda$, (i.e. the sky-projected angle between the stellar spin axis and planetary orbital plane). If the planet occults multiple latitudes, we can determine the stellar inclination (by disentangling it from $v \sin i$). Additionally, if you know the stellar rotation period (P$_{\rm{rot}}$) then you can use this in combination with  $v \sin i$ to determine the stellar inclination $i_{*}$. Then we can use this to model for $\lambda$ and determine the 3D obliquity, $\psi$. Here, $\psi$ is of great importance as it can provide insights into planetary migration/evolution and avoids biases introduced from only having $\lambda$. 

Additionally, the isolated starlight can be modelled to account for both the stellar rotation behind the planet and any CLV. Sun-like stars which possess a convective envelope have surfaces covered in granules, bubbles of hot plasma which rise to the surface (blueshift), before cooling and falling back into intergranular lanes (redshift). The net convective velocity shift caused by these granules changes as a function of limb angle (i.e. from the centre to the limb of the star) due to line-of-sight changes and the corrugated surface of the star. For example, towards the limb the granular walls become visible (due to geometrical effects) and the bottom and tops of the granules become less visible. Depending on the stellar type, this can cause the net convective blueshift (CB) to be less than compared to disk centre, where the walls of the granules are less visible and dimmer than the top and bottoms. This centre-to-limb CB has been observed on the Sun with RVs changing on the level of 100ms$^{-1}$ \citep[see][]{dravins1982photospheric} and on other stars including the K-dwarf HD~189733 \citep[see][]{czesla2015center, cegla2016rossiter}. Overall, these velocity shifts can impact the RM effect which is used to determine the projected obliquity \citep{cegla2016modeling}. Therefore, ignoring the velocity contributions caused by granulation can bias or skew these measurements and impact our understanding of planet formation and evolution. In addition to the RRM, other techniques utilising transiting exoplanets have been used to probe, detect and map the presence of starspots on the surface of stellar hosts \citep[see e.g.][and references therein]{silva2003method, wolter2009transit, sanchis2011starspots, morris2017starspots, movcnik2017recurring, zaleski2020activity}.

In our Solar System, all eight planets orbit the Sun in line with its rotation and are nearly aligned, with obliquities $\sim$7$^{\circ}$), suggesting all of the planets formed within a protoplanetary disk. Other star-planet systems have been found to be well-aligned and overall $\sim$ 120 of star-planet systems with measured obliquites are aligned. For example \cite{knudstrup2022orbital} measure $\lambda$ = -2$\pm$6$^{\circ}$ in the HD~332231 system (a warm giant planet orbiting an F8 star, P$_{\rm{orb}}$ = 18.7 d). They argue that measurements of obliquities are more meaningful for long period planets as a result of tidal interactions between the star and planet changing the star-planet alignment. \cite{christian2022possible} investigate the effects of wide visual binary companions on the formation and evolution of exoplanets. They find that planets to these binary systems tend to be aligned with the binary, demonstrating the alignment is not only due to the formation of the planet. There have been misaligned systems found with obliquities greater than 30$^{\circ}$ such as WASP-8~b \citep{queloz2010wasp, bourrier2017refined} and WASP-127~b \citep{allart2020wasp} in retrograde orbits ($\lambda \sim 180^{\circ}$; approximately 11 known), and WASP-79~b \citep{brown2017rossiter} with a polar orbit ($\lambda \sim 90^{\circ}$; approximately 8 known). It has been suggested these misaligned orbits could be a result of Kozai migration \citep[e.g. GJ~436~b][]{bourrier2018orbital} where the orbit of a planet is disturbed by a second body causing the argument of pericentre to oscillate around a constant value resulting in an exchange between its eccentricity and inclination, coupled with tidal friction \citep[e.g.][]{fabrycky2007shrinking,nagasawa2008formation}. Additionally, other mechanisms include Secular resonance crossing to explain the polar orbits of Neptune mass planets, magnetic warping which can cause young protoplanetary disks to tilt towards a perpendicular orientation and high eccentricity tidal migration which is often used to explain hot Jupiters \citep[see][for more details]{albrecht2021preponderance}. Therefore, classifying the obliquities of planetary systems is a vital tool for understanding how planets arrive in their orbits and what their fates will be over time \citep[see the review by][]{albrecht2022stellar}.

For this study we focus on the WASP-166~b system which is a transiting super-Neptune, within the Neptune desert, with a mass of $M_{\rm{p}} = 0.101 \pm 0.005$~M$_{\rm{Jup}}$ and radius $R_{\rm{p}} = 0.63 \pm 0.03$~R$_{\rm{Jup}}$ \citep{hellier2019wasp}. The star is a bright, $V = 9.36$, F9 main sequence dwarf with an age of $2.1 \pm 0.9$~Gyr. It was first discovered by WASP-South and later followed up by HARPS and CORALIE with a total of 220 and 41 RVs respectively \citep[see][for the full analysis]{hellier2019wasp}. \citet{hellier2019wasp} analysed RM observations with the HARPS spectrograph and found WASP-166~b to be aligned with the stellar rotation axis with $\lambda = 3 \pm 5^{\circ}$. In another study, \citet{vedad2021wasps} used the same HARPS observations to re-analyse the system using the RRM technique. They also found the system was aligned with $\lambda = 1.3\substack{+2.6 \\ -2.0}^{\circ}$. WASP-166 has also been central to various other studies: \citet{bryant20multicam} use simultaneous photometric transit observations of WASP-166~b from {\tess} and NGTS and found the multi-scope precision of NGTS can match {\tess}. Finally, \citet{seidel2020wasp166} use HARPS and then ESPRESSO \citep[][same observations as in this paper]{seidel2022hot} observations of WASP-166~b to detect neutral sodium in the planet atmosphere at a 3.4$\sigma$ level and tentative evidence of line broadening. This suggests winds are responsible for pushing sodium further into space causing the bloated nature of this highly irradiated world. 

\begin{table}
    \centering
    \caption{Summary of the ESPRESSO, TESS and NGTS data used in this work.}
    {\sl ESPRESSO}
    \begin{tabular}{ccccccc}
    \hline
    \hline
    Run & Night & $N_{\rm{obs}}$ & $t_{\rm{exp}}$ & $\gamma$$^a$ & SNR$^b$ & $\sigma_{\rm{RV}}$$^c$ \\
    & & & (s) & (kms$^{-1}$) & (550nm) & (cms$^{-1}$) \\
    \hline
    A & 31 Dec 2020 & 146 & 100 & 23.527 & 50 & 157 \\
    B & 18 Feb 2021 & 139 & 100 & 23.537 & 45 & 177\\
    \hline 
    \end{tabular}
    
    \vspace{5mm}
    {\sl TESS}
    \begin{tabular}{ccccc}
    \hline 
    \hline 
    Sector & Date & $N_{\rm{obs}}$ & $t_{\rm{exp}}$ & $\sigma_{\rm{residual}}$ \\
    & & & (s) & (ppm\,per\,2\,min) \\
    \hline
    8 & 2 Feb -- 27 Feb 2019 & 13045 & 120 & 565 \\
    35 & 9 Feb -- 6 Mar 2021 & 13686 & 120 & 600 \\
    \hline
    \end{tabular}
    
    \vspace{5mm}
    {\sl NGTS}
    \begin{tabular}{ccccc}
    \hline 
    \hline 
    No. Cameras & Date & $N_{\rm{obs}}$ & $t_{\rm{exp}}$ & $\sigma_{\rm{residual}}$ \\
    & & & (s) & (ppm\,per\,2\,min) \\
    \hline
    6 & 31 Dec 2020 & 9654 & 10 & 722 \\
    6 & 18 Feb 2021 & 13845 & 10 & 522 \\
    \hline 
    \end{tabular}
        \vspace{2mm}
     \begin{flushleft}
    {\bf Notes:} $^a$ The error on the systemic velocity, $\gamma$, is $\sim$0.017~kms$^{-1}$ for each run. $^b$ The SNR was computed as the average of order 112 where 550~nm falls on. $^c$ $\sigma$ stands for the average uncertainty of the RVs. 
    \end{flushleft}
    \label{observations}
\end{table}

In this paper we apply the RRM technique on newly acquired ESPRESSO observations of the WASP-166~b system to spatially resolve the stellar surface, by subtracting in-transit spectra from out-of-transit spectra, to characterise stellar differential rotation and centre-to-limb convection-induced variations, and to determine the star-planet obliquity. In \S \ref{sec:obs} we detail all of the photometric and spectroscopic observations used in this study. \S \ref{sec:trans_analysis} then details the transit analysis of the {\tess} sectors 8 and 35 and new NGTS photometric lightcurves where the transit parameters of the system are derived. Finally, we discuss the RRM analysis and results in \S \ref{sec:rrm} and \S \ref{sec:results} respectively. 

\section{Observations}\label{sec:obs}
For the analysis of the WASP-166 system we use both photometric data from {\tess} and NGTS as well as spectroscopic data from ESPRESSO. Here we give details of the observations and data reduction in more detail, where a summary of the observations can be found in Table \ref{observations}.

\begin{figure*}
    \centering
    \includegraphics[width=0.97\textwidth]{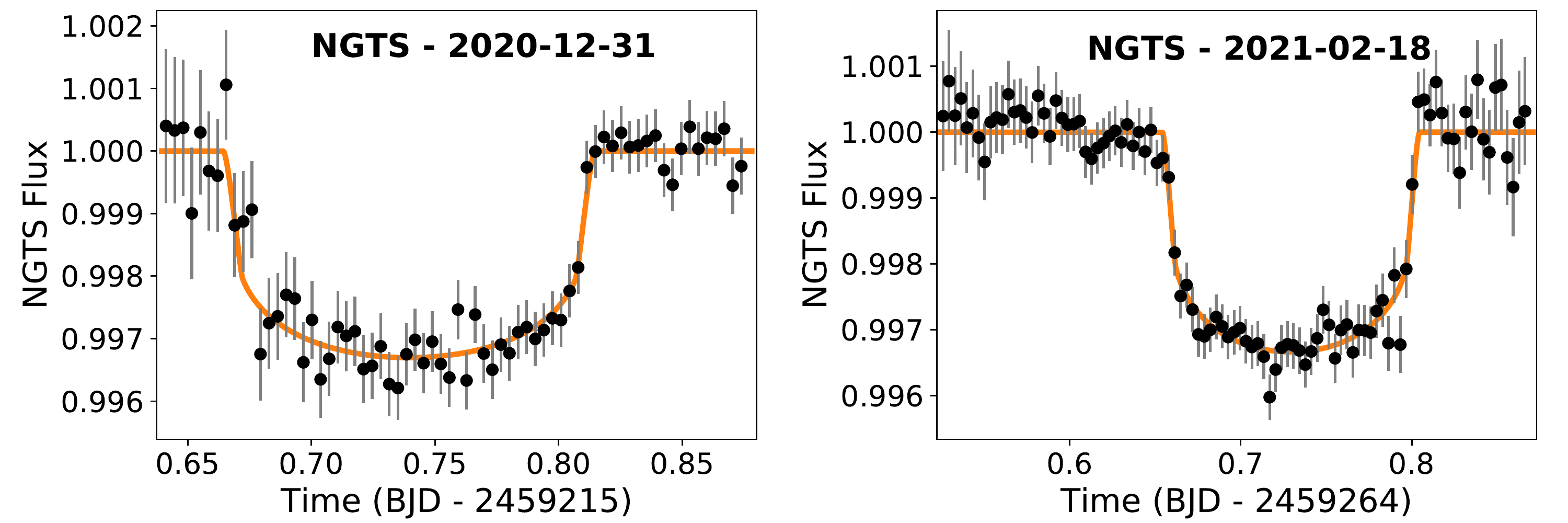}
    \caption{NGTS photometry of the two WASP-166~b transits from the nights of 2020 December 31 (left) and 2021 February 18 (right) which are simultaneous with the ESPRESSO observations. For both panels the black points give the detrended photometry from all six NGTS telescopes binned together on a timescale of 5\,minutes. The orange lines give the transit models calculated using the median parameters from the fits to the NGTS data (see Section~\ref{sec:trans_analysis}). There is no evidence of contamination during these transits from stellar activity.}
    \label{fig:ngts}
\end{figure*}

\subsection{Photometric Data}
To isolate the stellar spectrum behind the planet, in-transit spectra are directly subtracted from out-of-transit spectra. This is made possible by scaling the CCFs using photometric lightcurves. As a result, for WASP-166~b we utilise six {\tess} and two NGTS photometric transits to determine the transit, planet properties, and to scale the ESPRESSO CCFs. 

\subsubsection{TESS Photometry}\label{sec:tessphot}
WASP-166 was observed by the Transiting Exoplanet Survey Satellite \citep[{\tess}:][]{ricker2014tess} in 2-min cadence capturing a total of seven transits during Sector 8 and Sector 35 between the 2 to 27 February 2019 and 9 February to 6 March 2021, respectively. We accessed the 2-min lightcurves produced by the TESS Science Processing Operations Centre (SPOC) pipeline \citep{jenkins2016spoc} and use the \textsc{PDCSAP\_FLUX} time series for our analysis.

We used the transit ephemerides from \citet{hellier2019wasp} to mask out the in-transit data points and then ran a Lomb-Scargle analysis to search for and determine a stellar rotation period. Our analysis yielded no significant periodic signals. From a visual inspection of the light curve it is clear that there is some low level out-of-transit variability (see Figure~\ref{fig:tess_LC}); however, it is unclear whether this is astrophysical or instrumental in nature. To prevent this variability affecting our transit fit, we flatten the light curve using a spline fit, again masking out the in-transit data points to avoid the spline affecting the transit shapes. The methodology used may underestimate the errors on the planetary properties but this does not affect the RRM analyis in this work. The {\tess} photometry and detrending spline are shown in Figure~\ref{fig:tess_LC}. 

\subsubsection{NGTS Photometry}\label{sec:ngtsphot}
The Next Generation Transit Survey \citep[NGTS; ][]{wheatley2018ngts} is a photometric facility consisting of twelve independently operated robotic telescopes at ESO's Paranal Observatory in Chile. This is the same location as ESPRESSO, therefore, both instruments experience the same weather conditions. Each NGTS telescope has a 20\,cm diameter aperture and observes using a custom NGTS filter (520 -- 890\,nm).

We observed WASP-166 using NGTS simultaneously with the ESPRESSO transit observations on the nights of 31st December 2020 and 18th February 2021 (see Figure \ref{fig:ngts}). For both NGTS observations, we utilised six NGTS telescopes simultaneously to observe the transit event in the multi-telescope observing mode described in \citet{smith2020shallow} and \citet{bryant20multicam}. For both nights, an exposure time of $10\,$seconds was used and the star was observed at airmass $<$ 2. On the night of 31st December 2020 a total of 9654 images were taken across the six telescopes, and 13845 were taken on the night of 18th February 2021. 

The NGTS images were reduced using a version of the standard NGTS photometry pipeline, described in \citet{wheatley2018ngts}, which has been adapted for targeted single star observations. In short, standard aperture photometry routines are performed on the images using the SEP Python library \citep{bertin1996sextractor, barbary16sep}. For the WASP-166 observations presented in this work, we used circular apertures with a radius of five~pixels (25\,\arcsec). During this reduction comparison stars which are similar in magnitude and CCD position to the target star are automatically identified using the \textit{Gaia} DR2 \citep{GAIA, GAIA_DR2} and parameters found in the {\tess} input catalogue \citep[v8;][]{stassun2019ticv8}. Each comparison star selected is also checked to ensure it is isolated from other stars. Faint nearby stars within the aperture can contaminate the photometry and make the observed transit appear shallower than it is in reality. We queried the \textit{Gaia} DR2 \citep{GAIA_DR2} and found four faint stars ($G = 18.61 -- 19.87$\,mag) within 25\,\arcsec of WASP-166. Based on these neighbours we estimate a maximum possible contamination factor of just 0.049\%. Therefore, we expect the contamination from these faint stars to not have a significant impact on our results and so, we treat the NGTS data as undiluted throughout the analysis in this paper.

\subsection{Spectroscopic Data}\label{sec:espressospect}
Two transits of WASP-166~b were observed on the nights of 31st December 2020 (run A) and 18th February 2021 (run B) using the ESPRESSO \citep{pepe2014espresso, pepe2021espresso} spectrograph (380 - 788~nm) mounted on the Very Large Telescope (VLT) at the ESO Paranal Observatory in Chile (ID: 106.21EM, PI: H.M. Cegla). The ESPRESSO observations were carried out using UT1 for the first night and UT4 for the second under good conditions, with airmass varying between 1.0 -- 2.5'' and 1.0 -- 1.7'' for each run A and run B, respectively, in the high resolution mode using 2 $\times$ 1 binning. Exposure times were fixed at 100~s for each night to reach a SNR near 50 at 550~nm (to be photon noise dominated) and to ensure a good temporal cadence, with a 45~s dead-time per observation due to read out. Each run covered a duration of 6h 40m and 6h 15m of uninterrupted sequences covering the full transit duration and includes  2 -- 3~h of pre and post baseline. A summary of the ESPRESSO observations can be found in Table \ref{observations}. 

The spectra were reduced with version 2.2.8 of the ESPRESSO data reduction software (DRS), using a F9 binary mask to create cross-correlation functions (CCFs) which we use for our analysis. Additionally, the DRS also outputs the contrast (i.e. depth), full width at half max (FWHM: i.e width) and centroid (i.e. radial velocity) of each CCF. In run A the SNR can be seen to increase over the duration of the night which correlates with the decreasing airmass as observing conditions improve. However, in run B the SNR decreases towards the end of the night as a result of passing thin clouds. Despite this, the FWHM and contrast remain steady during both runs and are dispersed around the mean. Overall, the median integrated radial velocity uncertainties for run A and run B are 1.57 ms$^{-1}$ and 1.77 ms$^{-1}$, respectively. 

\section{Photometric Transit Analysis}\label{sec:trans_analysis}
When computing the RRM we spatially resolve the stellar surface by subtracting in-transit spectra from a master out-of-transit spectra. To do this, the spectra are scaled using a modelled transit lightcurve to remove any variations caused by the Earth’s atmosphere. This enables a direct subtraction between the spectra and avoids any assumptions on the local stellar line profile shapes. As a result, we need good knowledge of the transit parameters of the system, which we obtain from the spline detrended {\tess} photometry as the {\tess} data are more precise for this analysis. We identified that one of the {\tess} transits falls close to the start of the second orbit of Sector 8. {\tess} photometry at the start of orbits often suffers from increased systematic trends. Therefore, to avoid these trends affecting our analysis through this transit, we exclude it from the analysis.

\begin{figure}
    \centering
    \includegraphics[width=0.47\textwidth]{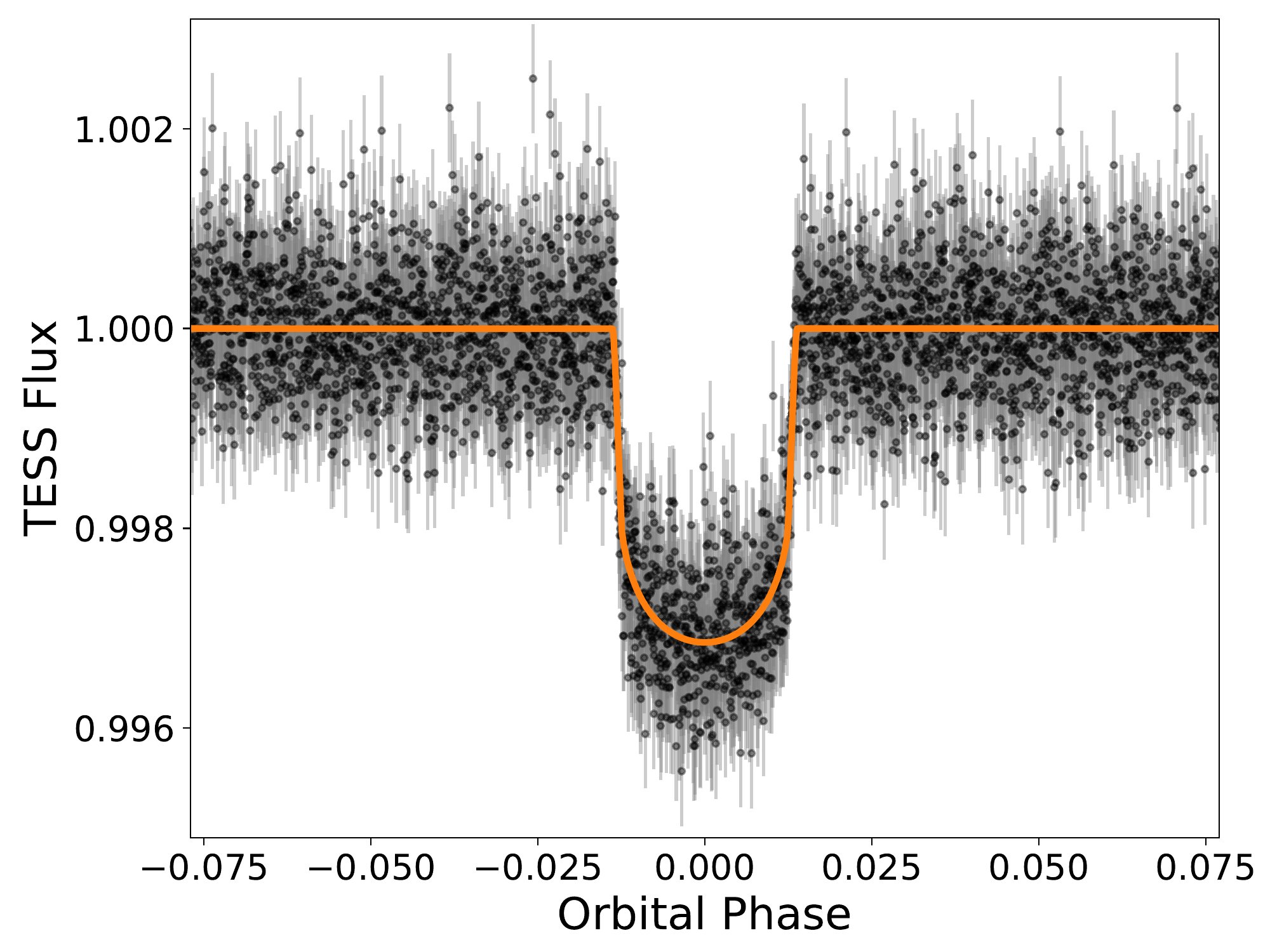}
    \caption{{\tess} lightcurves of six WASP-166~b transits from both Sectors 8 and 35 which have been flattened and phase-folded using the median orbital period derived from the transit modelling (Section~\ref{sec:trans_analysis}). The orange line gives the transit model obtained using the median parameters which are given in Table~\ref{planetary_properties}.}
    \label{fig:tess_phasefold}
\end{figure}

We fit the {\tess} photometry (600 - 1040~nm) using the MCMC method from the \textsc{emcee} package \citep{foremanmackey13} and the \textsc{batman} transit model \citep{kreidberg2015PASPbatman}. We take as free parameters: the time of transit centre ($T_C$), the orbital period (\Porb), the planet-to-star radius ratio ($R_{\rm{p}} / R_{\rm{*}}$), the scaled semi-major axis (a/$R_{\rm{*}}$), the orbital inclination ($i_{\rm{p}}$), the limb darkening coefficients ($u_1$ and $u_2$), and a constant flux offset ($f_0$). For the limb darkening, we use a quadratic law and calculate values for the coefficients using \textsc{LDTK} \citep{parviainan2015ldtk} and the stellar parameters from \citet{hellier2019wasp}. We calculate values of $u_{1, \ \rm LDTK} = \Nuonetsprior$ and $u_{2, \ \rm LDTK} = \Nutwotsprior$, which we adopt as Gaussian priors for the fitting. For the remaining parameters we adopt the \citet{hellier2019wasp} values as initial positions for the walkers and uninformative priors that ensure a physically realistic solution. We run the MCMC using 20 walkers for 10000 steps per walker, following a burn in processes of 3000 steps. The transit model obtained from this fitting is displayed in Figure~\ref{fig:tess_phasefold} and the transit parameters are presented in Table~\ref{planetary_properties}.

\begin{table}
    \centering
    \caption{Planetary system properties for WASP-166. Fixed parameters were taken from \citet{hellier2019wasp}.}
    \begin{tabular}{l|c}
    \hline 
    Planetary Parameters & Value \\
    \hline
    {\bf Fitted Parameters} & \\
    \hline
    $T_{\rm{0}}$ (BJD) & \Ntc \\
    $P_{\rm{orb}}$ (days) & \Nperiod \\
    $R_{\rm{p}} / R_{\rm{*}}$ & \Nrprs \\
    a/$R_{\rm{*}}$ & \Nars \\
    $i_{\rm{p}}$ ($^{\circ}$) & \Ninc \\
    $u_1$ & \Nuonets \\
    $u_2$ & \Nutwots \\
    $f_0$ & \Nfzero \\
    \hline
    {\bf Derived Parameters} & \\
    \hline 
    $T_{\rm{dur}}$ (hours) & \Nduration \\ 
    a (AU) & \Naau \\
    Impact Parameter ($b$) & \Nimpact \\
    $R_{\rm{p}}$ ($R_{\rm{Jup}}$) & \Nrpl \\
    \hline 
    {\bf Fixed Parameters} & \\
    \hline 
    $e$ & 0.0 \\
    $\omega$ ($^{\circ}$) & 90.0 \\
    K (m/s) & 10.4  \\
    $T_{\rm{eff}}$ (K) & 6050 \\
    log $g$ & 4.5 \\
    \hline 
    \end{tabular}
    \label{planetary_properties}
\end{table}

The NGTS transits (520 - 890~nm) are then individually fit to investigate whether the transits are significantly early or late compared to the prediction from the {\tess} model. This is a vital step for computing the RRM technique as we need the most precise ephemeris possible. For these NGTS fits, we produce a transit model using the {\tess} fit parameters, and take $T_C$, $R_{\rm{p}} / R_{\rm{*}}$, and the limb darkening coefficients as the only free parameters. We compute new limb darkening coefficients of $u_{1, \ \rm LDTK} = \Nuonengprior$ and $u_{2, \ \rm LDTK} = \Nutwongprior$ for the NGTS band pass, which we use as Gaussian priors. We also impose a Gaussian prior on $R_{\rm{p}} / R_{\rm{*}}$ of $0.052 \pm 0.001$ based on the results from the fit to the {\tess} data.

During the NGTS fitting, we detrend the six independent telescope light curves for each night. We do this by fitting a linear out-of-transit model with time to each light curve simultaneously with the transit model. From these fits, we measure NGTS transit times of $T_{\rm C, NGTS-1} = \Ntcngone$ and $T_{\rm C, NGTS-2} = \Ntcngtwo$. These transit times are consistent with the {\tess} predictions at the levels of 0.83\,$\sigma$ and 1.47\,$\sigma$. Therefore, we do not have any significant indication of timing variations for the two transit events observed with ESPRESSO.

However, slight inaccuracies in the ephemeris at the time of the ESPRESSO observations can bias the RRM analysis \citep{kunovac2021orbital}. Therefore, we fit all the photometric data together to derive the most accurate ephemeris for WASP-166~b. We use the flattened {\tess} and the two detrended six telescope NGTS observations, as well as the nine telescope NGTS transit observation of WASP-166~b from \citet{bryant20multicam}. For this final fit, we use the parameters derived from the {\tess} analysis, and allow $T_C$ and \Porb\ to vary. We find values of $T_C$ = 2458524.40869 $\pm$ 0.00030 and \Porb\,= 5.443542$ \pm$ 0.000003\,days. The new ephemeris (determined from  both NGTS and {\tess} transits) along with the planet-to-star radius ratio ($R_{\rm{p}} / R_{\rm{*}}$), the scaled semi-major axis (a/$R_{\rm{*}}$), the orbital inclination, ($i_{\rm{p}}$) and the limb darkening coefficients ($u_1$ and $u_2$) (from the {\tess} alone transit fitting) are detailed in Table \ref{planetary_properties} as the final planetary properties used in the RRM analysis. 

In Section~\ref{sec:ngtsphot}, we mentioned that we expected any dilution of the transit depth due to contamination of the NGTS aperture to not have a significant effect on our results. From this analysis, we derive a transit depth of $3137.6\pm 23.8$\,ppm. Assuming the contamination factor of 0.049\% obtained in Section~\ref{sec:ngtsphot}, we calculate an ``undiluted'' depth of $3139.1\pm 23.8$\,ppm. The difference between these two depths is just $1.55\pm 33.62$\,ppm, clearly confirming that any possible effect had by contamination from the faint stars within the NGTS aperture on the results of this study will be insignificant.

The differences in bandpass between ESPRESSO, NGTS and {\tess} means there will be slight changes in the limb darkening and $R_{\rm{p}} / R_{\rm{*}}$. Therefore, we tested the impact these differences would have on the local RVs (see Figure \ref{local_RVs}) using the LDTK {\tt python} package to determine limb darkening parameters in the ESPRESSO bandpass and using the $R_{\rm{p}} / R_{\rm{*}}$ determined in the NGTS bandpass from \citet{bryant20multicam}. Overall, we found there was no significant change in the local RVs and any change was within 1-sigma. Additionally, we checked the impact of any inaccuracies in the ephemeris on the local RVs by using the upper and lower limits of the errors on T$_0$ and P$_{\rm{rot}}$ from Table \ref{planetary_properties}; we found this did not have a significant change in the RRM analysis. Finally, the simultaneous NGTS transits are used to ensure there has been no contamination with stellar activity during our ESPRESSO observations.

\begin{figure}
    \centering
    \includegraphics[width = 0.47\textwidth]{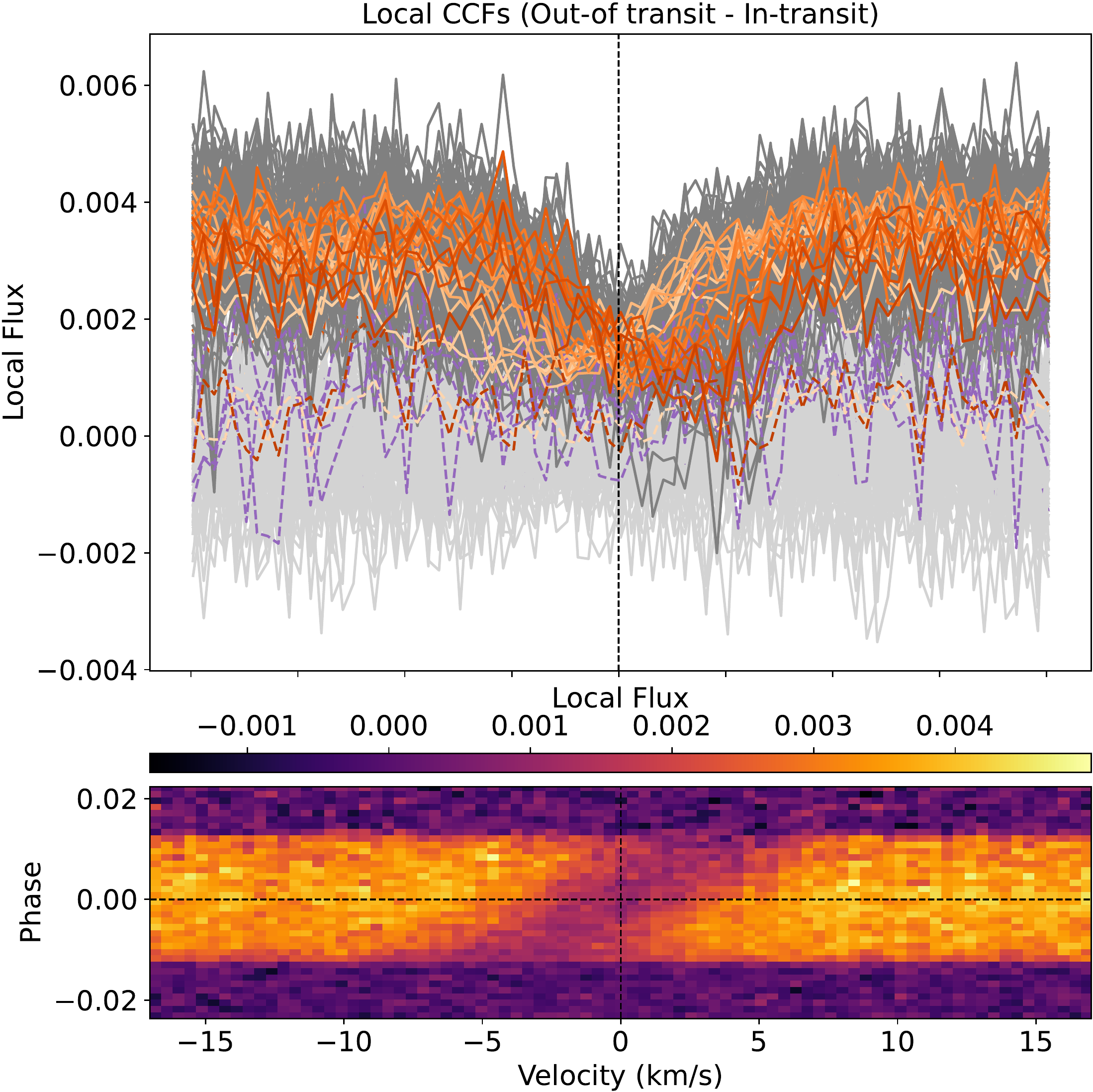}
    \caption{{\it Top:} The local CCFs (out-of-transit – in-transit CCFs) of the star behind the transiting super-Neptune WASP-166 b. The dark grey lines represent the un-binned in-transit observations, the light grey are the out-of-transit observations and the orange lines are the binned 10 minute in-transit observations. The changing gradient of the orange lines represents the changing centroid position where the darker orange is more redshifted. Dashed lines are observations for binned (orange) and un-binned (purple) which have a stellar disk position $\langle\mu\rangle <$ 0.25 and are not used in the analysis. A dotted line at 0~kms$^{-1}$ is included to guide the eye. {\it Bottom:} A top view of the top plot showing a map of the time series local CCFs colour coded by the local flux. A dotted line at phase zero and 0~kms$^{-1}$ are included to guide the eye. As the planet crosses the stellar disk a streak can be seen in the map. Transit centre occurs at 0~kms$^{-1}$ in both plots because the Keplerian stellar motion and systemic RVs were removed. }
    \label{local_ccfs}
\end{figure}

\section{The Reloaded Rossiter McLaughlin Analysis}\label{sec:rrm}
To isolate the starlight behind the planet during its transit, we use the reloaded RM (RRM) technique. From here on we will use the term local CCF (\locccf) to refer to the occulted light emitted behind the planet and the term disk-integrated CCF (\DIccf) to refer to the light emitted by the entire stellar disk. Here we will discuss the RRM technique at a high level, therefore, we refer the reader to \cite{cegla2016rossiter} for further details.

\subsection{Isolating the Planet Occulted Starlight}
Firstly, the \DIccfs from the ESPRESSO observations are shifted and re-binned in velocity space to correct for the Keplerian motions of the star induced by WASP-166~b, using the orbital properties in Table \ref{planetary_properties}. Using the ephemeris and transit duration from Table \ref{planetary_properties}, we separated the \DIccfs into those in-transit and those out-of-transit. We then created a single master-out \DIccf for each night by summing all out-of-transit \DIccfs and normalising the continuum to unity. By fitting the master-out \DIccf for each night with a Gaussian profile we can determine the systemic velocity, $\gamma$, as the centroid, see Table \ref{observations}. We define the continuum as regions which lie outside the region of $\pm$ 10~kms$^{-1}$, a combination of the $v_{\rm{eq}}\sin i_*$ and FWHM of the \locccfs, from the stars rest velocity (i.e. $\gamma$ has been subtracted). Before any further analysis, all \DIccfs were shifted to the stellar rest frame by subtracting $\gamma$ for each corresponding night from the x-axis (i.e. there was no re-binning). As the ESPRESSO observations are not calibrated photometrically, we have to continuum scale each \DIccf to reflect the change in flux absorbed by the planetary disk. To do this, we normalise each \DIccf by their individual continuum value and then scale them using a quadratic limb darkened transit model using the fitted {\tess} parameters in Table \ref{planetary_properties}. Finally, we can determine the \locccfs which represent the CCF of the occulted starlight behind the planet by subtracting the now scaled in-transit \DIccfs from the master-out \DIccf, see Figure \ref{local_ccfs}.  

\begin{figure*}
    \centering
    \includegraphics[width = 0.97\textwidth]{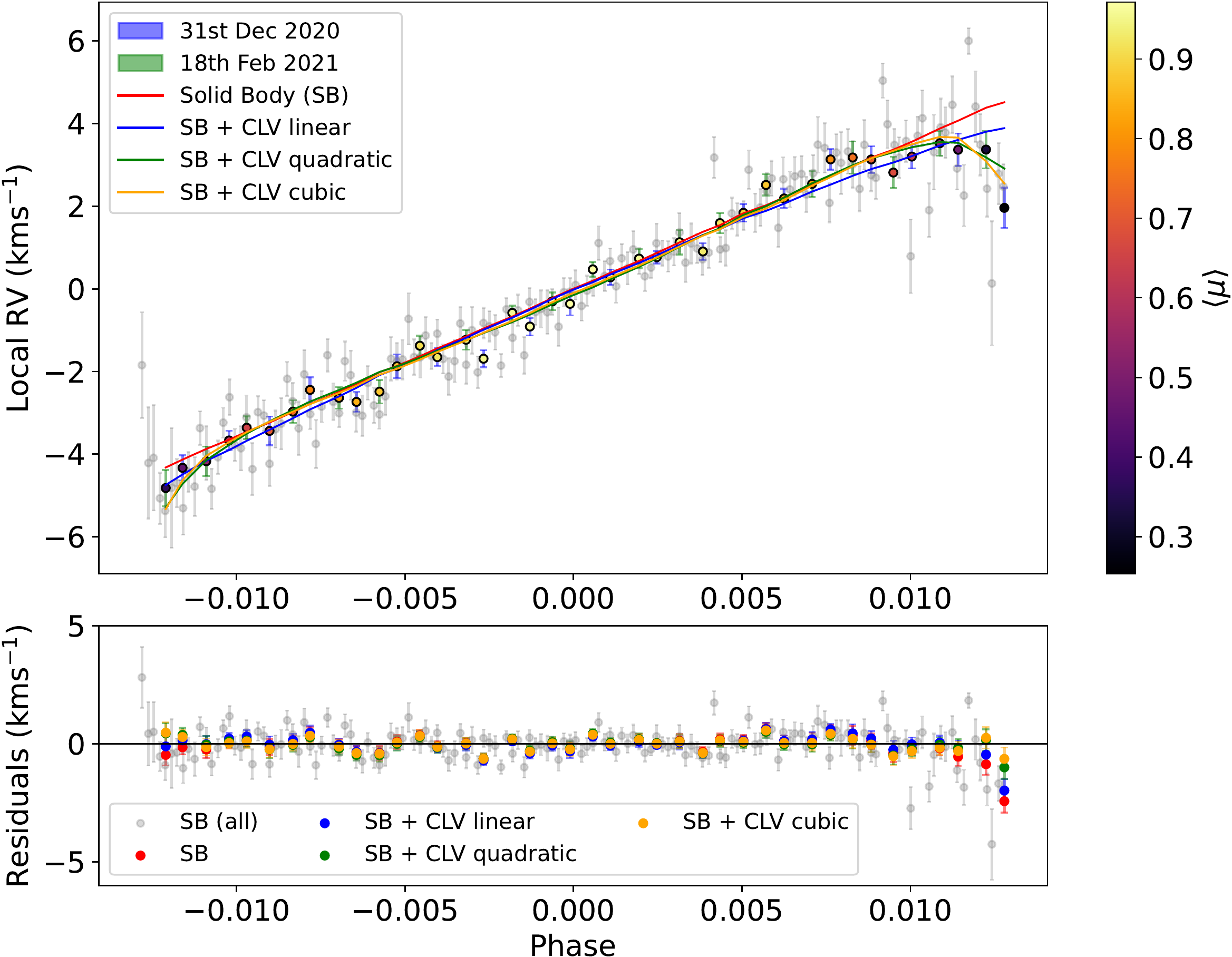}
    \caption{The top panel shows the local RVs for the un-binned observations (grey) and for ten minute binned data (colour), determined from the local CCFs of the regions occulted by the planet as a function of phase. The ten minute binned data points are colour coded by the stellar disk position behind the planet in units of brightness weighted $\langle\mu\rangle$ (where $\mu = \cos\theta$). The best-fit model for solid body rotation (SB) for the ten minute binned data (red line) is shown, along with the SB plus centre-to-limb linear (blue), SB plus centre-to-limb quadratic (green) and SB plus centre-to-limb cubic (orange) models. The bottom panel is the residuals (local RVs - model) for all models with colours corresponding to the top panel model lines, with a horizontal line at 0 to guide the eye. Points close to the limb can suffer from systematics (originating from the ephemeris, limb darkening etc.), therefore, we reran the model fitting removing the last point (Phase = 0.012) and found our results were within 1$\sigma$ with the BIC favouring the SB plus centre-to-limb cubic model.}
    \label{local_RVs}
\end{figure*}

\subsection{Determining and Modelling the Stellar Velocity of the Planet Occulted Starlight}\label{get_vel}
The \locccf (both binned and un-binned) in Figure \ref{local_ccfs} represent the starlight behind the planet along the transit chord, meaning they are both spectrally and spatially resolved. There is a change in continuum flux for \locccfs which are near the limb of the stellar disk due to limb darkening and partial occultation of the stellar disk by the planet during ingress and egress. To determine the stellar velocity of the occulted starlight, we fit Gaussian profiles using {\tt curve\_fit} from the {\tt Python Scipy} package \citep{2020SciPy-NMeth} to each of the \locccfs. There are a total of four Gaussian parameters in our fit including the offset (i.e. continuum), amplitude, centroid, and width. Initial values for these parameters for {\tt curve\_fit} were determined as continuum minus the minimum point of flux for the amplitude, the minimum flux point in velocity space between $\pm$ 10~kms$^{-1}$ for the centroid, the continuum for the offset, and the variance of the data for the width. The continuum was determined as any points which lie outside $\pm$ 10~kms$^{-1}$ and then lie between $\pm$ 0.5$\sigma$ of the mean. We applied constraints to both the centroid and offset during the fitting where the centroid was to be between $\pm$ 10~kms$^{-1}$ (a combination of the $v\sin i$ and FWHM of the \locccfs) and the offset was allowed to vary between $\pm$ 0.5$\sigma$ of the the mean of points which lie outside $\pm$ 10~kms$^{-1}$. The flux errors assigned to each \locccf were propagated from the errors on each \DIccf as determined from the version 2.2.8 of the ESPRESSO DRS. The local RVs of the planet occulted starlight are shown in Figure \ref{local_RVs}, plotted as a function of both phase and stellar disk position behind the planet. Cool main sequence stars show solar-like oscillations, known as pressure modes (p-modes), which are caused by sound waves travelling through the stellar interior. P-mode oscillations are present within the local RVs of WASP-166, therefore, we look to binning them out, see Section \ref{clv}.We removed CCFs with $\langle\mu\rangle < 0.25$ from our analysis, as profiles close to the limb were very noisy and when comparing the depth to the noise the signal was not significant to enable a Gaussian fit, see Figure \ref{local_ccfs} where they are shown as dashed lines. This resulted in 3 CCFs being removed from the binned 10 minute data analysis and 11 from the un-binned observations.

We fit the local RVs in Figure \ref{local_RVs} using the model by \cite{cegla2016rossiter} (hereafter referred to as the RRM model). This RRM model consists of one key formula (see equation 10 of \citet{cegla2016modeling}) which depends on the position of the transiting planet centre with respect to the stellar disk, the sky-projected obliquity ($\lambda$), the stellar inclination ($i_*$), the equatorial rotational velocity ($v_{eq}$), the differential rotational shear ($\alpha$), quadratic stellar limb darkening (u$_1$ and u$_2$), and centre-to-limb convective variations of the star ($v_{\rm{conv}}$, see equation 11). We use the same coordinate system and angle conventions as used in Figure 3 of \cite{cegla2016rossiter}. The stellar inclination is the angle from the line-of-sight towards the stellar spin axis (between 0 -- 180$^{\circ}$) and $\lambda$ is the sky-projected angle between the stellar spin axis and planetary orbital plane(between -180 -- 180$^{\circ}$). For WASP-166, we use different scenarios for the RRM model depending on whether or not we account for differential rotation and centre-to-limb convective variations. For every ESPRESSO observation, we calculate the brightness weighted rotational velocity behind the planet following \citet{cegla2016modeling} and using the quadratic limb darkening coefficients from the Table \ref{planetary_properties}.

We sample the RRM model parameters using a Markov Chain Monte Carlo (MCMC) method implemented in the {\tt Python} package {\tt emcee} \citep{foremanmackey13}. We used a total of 200 walkers with 2000 steps and a burn in phase of 200 steps which was discarded. We set uniform priors on all the RRM model parameters and let $v_{eq}\sin i_* \sim \mathcal{U}(0, 100) km^{-1}$ and $\lambda \sim \mathcal{U}(-180^{\circ}, 180^{\circ})$ for the solid body scenario. When including differential rotation we let $\alpha \sim \mathcal{U}(-1.0, 1.0)$, $i_* \sim \mathcal{U}(0^{\circ}, 180^{\circ})$, $v_{eq} \sim \mathcal{U}(0, 100) km^{-1}$ and $\lambda \sim \mathcal{U}(-180^{\circ}, 180^{\circ})$. We initiate the walkers in a tiny Gaussian ball around the maximum likelihood result and display the final RRM model parameters in Table \ref{mcmc_fits} with uncertainties based on the 16th, 50th, and 84th percentiles. 

\section{Results}\label{sec:results}

\begin{table*}
    \centering
        \caption{MCMC observational results for WASP-166 and the derived 3D spin-orbit obliquity.}
    \resizebox{1.0\textwidth}{!}{    
    \begin{tabular}{lccccccccccccc}
    \hline
        Model & No. of Model & $v_{\rm{eq}}$ & $i_{\rm{*}}$ & $\alpha$ & $\lambda$    & $\sigma$     & $c_1$        & $c_2$        & $c_3$        & BIC & \chisq &  $\chi_{\nu}^{2}$ &  $\psi$  \\
              & Parameters   & (kms$^{-1}$)  & ($^{\circ}$) &          & ($^{\circ}$) & (kms$^{-1}$) & (kms$^{-1}$) & (kms$^{-1}$) & (kms$^{-1}$) &     &       &  &($^{\circ}$) \\
        \hline
        {\bf Un-binned observations} &&&&&&&&&&\\
        \hline
        SB & 2 &  $4.89\substack{+0.08 \\ -0.08}$ & 90.0 & 0.0 & $-4.49\substack{+1.74 \\ -1.73}$ & -- & -- & -- & -- & 311.34 & 301.32 & 2.01 &-- \\
        SB + $\sigma$ & 3 & $4.84\substack{+0.10 \\ -0.11}$ & 90.0 & 0.0 & $-3.93\substack{+2.57 \\ -2.57}$ & $0.43\substack{+0.06 \\ -0.05}$ & -- & -- & -- &  & & & -- \\
        SB + CLV1 & 3 & $4.89\substack{+0.08 \\ -0.08}$ & 90.0 & 0.0 & $-4.51\substack{+2.26 \\ -2.28}$ & -- & $0.01\substack{+0.22 \\ -0.21}$ & -- & -- & 316.35 & 301.32 & 2.01 & -- \\
        SB + CLV2 & 4 & $4.95\substack{+0.08 \\ -0.08}$ & 90.0 & 0.0 & $-6.49\substack{+2.27 \\ -2.25}$ & -- & $6.74\substack{+1.70 \\ -1.70}$ & $-4.70\substack{+1.18 \\ -1.18}$ & -- & 305.39 & 285.35 & 1.90 & -- \\
        SB + CLV3 & 5 & $4.98\substack{+0.08 \\ -0.08}$ & 90.0 & 0.0 & $-8.44\substack{+2.48 \\ -2.44}$ & -- & $21.56\substack{+7.82 \\ -7.80}$ & $-28.29\substack{+12.19 \\ -12.15}$ & $11.77\substack{+5.99 \\ -6.05}$ & 306.71 & 281.66 & 1.88 & -- \\
        DR (away) & 4 & $6.03\substack{+1.61 \\ -0.97}$ & $51.99\substack{+22.83 \\ -14.40}$ & $-0.33\substack{+0.21 \\ -0.31}$ & $-3.75\substack{+1.63 \\ -1.77}$ & -- & -- & -- & -- & 321.37 & 301.328 & 2.01 & $88.38\substack{+0.20 \\ -0.01}$ \\
        DR (towards) & 4 & $4.96\substack{+1.14 \\ -0.37}$ & $116.71\substack{+16.68 \\ -12.08}$ & $-0.40\substack{+0.21 \\ -0.29}$ & $-5.80\substack{+2.18 \\ -2.23}$ & -- & -- & -- & -- & 321.37 & 301.321 & 2.01 & $89.49\substack{+0.46 \\ -0.31}$ \\
        \hline
        {\bf Binned 10 minute observations} &&&&&&&&&&\\
        \hline
        SB & 2 &  $4.77\substack{+0.09 \\ -0.09}$ & 90.0 & 0.0 & $-5.93\substack{+1.99 \\ -2.01}$ & -- & -- & -- & -- & 86.74 & 80.62 & 1.98 & -- \\
        SB + $\sigma$ & 3 & $4.73\substack{+0.12 \\ -0.12}$ & 90.0 & 0.0 & $-7.05\substack{+2.84 \\ -2.94}$ & $0.04\substack{+0.23 \\ -0.31}$ & -- & -- & -- &  & & & -- \\
        SB + CLV1 & 3 & $4.83\substack{+0.10 \\ -0.09}$ & 90.0 & 0.0 & $-11.49\substack{+2.68 \\ -2.65}$ & -- & $0.76\substack{+0.26 \\ -0.25}$ & -- & -- & 80.62 & 69.55 & 1.73 & -- \\
        SB + CLV2 & 4 & $4.92\substack{+0.10 \\ -0.10}$ & 90.0 & 0.0 & $-13.95\substack{+2.68 \\ -2.62}$ & -- & $9.05\substack{+1.99 \\ -1.98}$ & $-5.75\substack{+1.36 \\ -1.38}$ & -- & 111.0 & 96.24 & 2.40 & -- \\
        SB + CLV3 & 5 & $4.97\substack{+0.11 \\ -0.11}$ & 90.0 & 0.0 & $-15.52\substack{+2.85 \\ -2.76}$ & -- & $22.26\substack{+9.16 \\ -9.15}$ & $-26.72\substack{+14.35 \\ -14.26}$ & $10.41\substack{+7.01 \\ -7.11}$ & 66.89 & 48.44 & 1.21 & -- \\
        DR + CLV1 (towards) & 5 & $5.82\substack{+0.53 \\ -0.40}$ & $133.07\substack{+6.41 \\ -5.70}$ & $-0.80\substack{+0.19 \\ -0.14}$ & $-30.04\substack{+6.26 \\ -4.48}$ & -- & $2.39\substack{+0.54 \\ -0.62}$ & -- & -- & 63.95 & 45.51 & 1.13 & $90.05\substack{+0.14 \\ -0.11}$ \\
        DR + CLV1 (away) & 5 & $7.78\substack{+2.56 \\ -1.44}$ & $33.35\substack{+10.09 \\ -9.55}$ & $-0.55\substack{+0.22 \\ -0.25}$ & $-7.93\substack{+2.13 \\ -2.35}$ & -- & $0.59\substack{+0.24 \\ -0.23}$ & -- & -- & 78.79 & 60.35 & 1.50 & $88.41\substack{+0.08 \\ -0.04}$ \\
        DR + CLV2 (towards) & 6 & $6.08\substack{+1.13 \\ -0.64}$ & $135.00\substack{+8.93 \\ -9.23}$ & $-0.50\substack{+0.19 \\ -0.23}$ & $-22.55\substack{+5.47 \\ -6.68}$ & -- & $6.24\substack{+2.33 \\ -2.37}$ & $-3.24\substack{+1.86 \\ -1.75}$ & -- & 154.10 & 131.97 & 3.29 & $90.06\substack{+0.22 \\ -0.21}$ \\
        DR + CLV2 (away) & 6 & $6.92\substack{+2.08 \\ -0.95}$ & $39.73\substack{+9.90 \\ -10.67}$ & $-0.57\substack{+0.23 \\ -0.26}$ & $-10.86\substack{+2.33 \\ -2.56}$ & -- & $9.24\substack{+2.18 \\ -2.21}$ & $-5.97\substack{+1.52 \\ -1.49}$ & -- & 105.33 & 82.80 & 2.07 & $88.39\substack{+0.06 \\ -0.01}$ \\
        DR + CLV3 (towards) & 7 & $6.10\substack{+0.09 \\ -0.57}$ & $133.64\substack{+8.42 \\ -7.98}$ & $-0.48\substack{+0.21 \\ -0.28}$ & $-23.78\substack{+6.12 \\ -8.08}$ & -- & $13.97\substack{+10.20 \\ -9.75}$ & $-15.41\substack{+15.13 \\ -15.49}$ & $6.04\substack{+7.65 \\ -7.45}$ & 67.86 & 42.00 & 1.05 & $90.03\substack{+0.20 \\ -0.16}$ \\
        DR + CLV3 (away) & 7 & $6.68\substack{+1.58 \\ -0.78}$ & $42.03\substack{+9.13 \\ -9.60}$ & $-0.58\substack{+0.25 \\ -0.26}$ & $-11.68\substack{+2.52 \\ -2.77}$ & -- & $16.72\substack{+9.58 \\ -9.79}$ & $-17.42\substack{+14.72 \\ -15.12}$ & $5.57\substack{+7.54 \\ -7.13}$ & 67.48 & 41.60 & 1.04 &  $88.39\substack{+0.003 \\ -0.04}$ \\
        \hline
        {\bf Binned 20 minute observations} &&&&&&&&&&\\
        \hline
        SB & 2 &  $4.80\substack{+0.10 \\ -0.10}$ & 90.0 & 0.0 & $-5.17\substack{+2.14 \\ -2.13}$ & -- & -- & -- & -- & 33.30 & 27.41 & 1.44 & -- \\
        SB + $\sigma$ & 3 & $4.78\substack{+0.12 \\ -0.13}$ & 90.0 & 0.0 & $-5.05\substack{+2.91 \\ -2.91}$ & $0.01\substack{+0.20 \\ -0.18}$ & -- & -- & -- &  & & & -- \\
        SB + CLV1 & 3 & $4.80\substack{+0.15 \\ -0.15}$ & 90.0 & 0.0 & $-3.81\substack{+4.67 \\ -4.68}$ & -- & $0.71\substack{+0.43 \\ -0.43}$ & -- & -- & 53.32 & 44.49 & 2.34 & -- \\
        SB + CLV2 & 4 & $4.81\substack{+0.10 \\ -0.10}$ & 90.0 & 0.0 & $-7.23\substack{+3.85 \\ -3.80}$ & -- & $3.70\substack{+2.85 \\ -2.82}$ & $-2.52\substack{+1.86 \\ -1.87}$ & -- & 40.09 & 28.32 & 1.49 & -- \\
        SB + CLV3 & 5 & $4.80\substack{+0.10 \\ -0.10}$ & 90.0 & 0.0 & $-5.79\substack{+4.49 \\ -4.51}$ & -- & $-3.36\substack{+11.48 \\ -11.34}$ & $8.27\substack{+16.83 \\ -17.12}$ & $-5.22\substack{+8.23 \\ -8.10}$  & 39.82 & 25.10 & 1.32 & -- \\
        \hline
        {\bf Line Core - Binned 10 minute observations} &&&&&&&&&&\\
        \hline
        SB & 2 &  $5.09\substack{+0.08 \\ -0.08}$ & 90.0 & 0.0 & $1.01\substack{+1.70 \\ -1.69}$ & -- & -- & -- & -- & 83.98 & 76.60 & 1.92 & -- \\
        SB + CLV1 & 3 & $5.16\substack{+0.09 \\ -0.09}$ & 90.0 & 0.0 & $4.78\substack{+2.41 \\ -2.51}$ & -- & $-0.51\substack{+0.25 \\ -0.25}$ & -- & -- & 83.53 & 72.46 & 1.81 & -- \\
        SB + CLV2 & 4 & $5.16\substack{+0.09 \\ -0.09}$ & 90.0 & 0.0 & $2.79\substack{+2.61 \\ -2.61}$ & -- & $4.25\substack{+2.12 \\ -2.11}$ & $-3.25\substack{+1.43 \\ -1.44}$ & -- & 82.06 & 67.31 & 1.68 & -- \\
        \hline
        {\bf Line Wing - Binned 10 minute observations} &&&&&&&&&&\\
        \hline
        SB & 2 &  $4.61\substack{+0.22 \\ -0.22}$ & 90.0 & 0.0 & $-9.31\substack{+4.24 \\ -4.16}$ & -- & -- & -- & -- & 49.18 & 41.95 & 1.13 & -- \\
        SB + CLV1 & 3 & $4.92\substack{+0.31 \\ -0.29}$ & 90.0 & 0.0 & $-23.80\substack{+6.37 \\ -5.90}$ & -- & $1.61\substack{+0.58 \\ -0.59}$ & -- & -- & 45.23 & 34.40 & 0.93 & -- \\
        SB + CLV2 & 4 & $4.92\substack{+0.31 \\ -0.29}$ & 90.0 & 0.0 & $-23.87\substack{+6.36 \\ -5.94}$ & -- & $6.64\substack{+4.79 \\ -4.76}$ & $-3.52\substack{+3.31 \\ -3.32}$ & -- & 47.72 & 33.28 & 0.90 & -- \\
        
        \hline
        
    \end{tabular}}
    \label{mcmc_results}
    \vspace{2mm}
     \begin{flushleft}
   {\bf Notes:} For all SB models $i_*$ and $\alpha$ are fixed under the assumption of rigid body rotation and the $v_{\rm{eq}}$ column corresponds to $v_{\rm{eq}}\sin i_*$. For these models we are unable to determine the 3D obliquity, $\psi$. The BIC of each model was calculated using \chisq, therefore, due to different sample sizes it is not possible to compare models between data sets. As a result, the reduced chi-squared ($\chi_{\nu}^{2}$) has been added to make these comparisons possible. For clarity, CLV1, CLV2 and CLV3 correspond to centre-to-limb linear, quadratic and cubic respectively. The first four corner plots for the un-binned observation MCMC runs are in an appendix and the remaining are available as supplementary material online.    
    \end{flushleft}
    \label{mcmc_fits}
\end{table*}

A visual inspection of the local RVs in Figure \ref{local_RVs} can provide an indication of the alignment of the WASP-166 system. Overall, the measured local RVs increase with orbital phase from negative to positive as the planet transits the stellar disk. Additionally, there is symmetry within the velocities with respect to phase, suggesting the system is in an aligned orbit with equal time spent crossing the blue and red-shifted regions. In two previous studies measurements for the projected obliquity, $\lambda$, and $v_{\rm{eq}}\sin i_*$ were obtained using HARPS data by \citet{hellier2019wasp} from spectroscopic line broadening and an MCMC fit to the RM effect ($\lambda = 3 \pm 5^{\circ}$ and $v_{\rm{eq}}\sin i_* = 5.1 \pm 0.3$~kms$^{-1}$) and RRM modelling by \citet{vedad2021wasps} ($\lambda = 1.26^{+2.94}_{-2.95}$$^{\circ}$ and $v_{\rm{eq}}\sin i_* = 4.82^{+0.23}_{-0.21}$~kms$^{-1}$). All of the results for our model fitting discussed in the following sections can be found in Table \ref{mcmc_fits}. 

\subsection{Solid Body Alone Model}
To analyse the local RVs further, we first applied the RRM model assuming a solid body (SB) rotation for the star as it has the least number of free parameters. As a result, this leads to a degeneracy between $v_{\rm{eq}}$ and $\sin i_*$, preventing us from calculating the stellar latitudes occulted by the planet. The two RRM model parameters are, therefore, the sky-projected obliquity, $\lambda$, and the projected rotational velocity, $v_{\rm{eq}}\sin i_*$. The probability distributions, from the MCMC model fitting, along with the 1D distributions are displayed as a corner plot in Appendix \ref{corner:SB}. We find, for the combined fit to both runs $v_{\rm{eq}}\sin i_* = 4.89 \pm 0.08$~kms$^{-1}$ and $\lambda = -4.49\substack{+1.74 \\ -1.73}^{\circ}$. These values are consistent with measurements obtained using HARPS data by \citet{hellier2019wasp} and \citet{vedad2021wasps}. The derived projected obliquity informs us that the system is in an aligned orbit as suspected from a visual inspection of the local RVs. The SB model fit to the data is shown in Figure \ref{local_RVs}. We used an Anderson Darling test to check if the residuals in Figure \ref{local_RVs} were normally distributed and found they were not indicating there is remaining correlated red noise, potentially as a result of magneto convection (i.e. granulation/super granulation) and or residual p-modes.

In addition to fitting both runs simultaneously, we also fit them individually to check their consistency. For run A we derive $\lambda = -9.60\substack{+2.36 \\ -2.37}^{\circ}$ and $v_{\rm{eq}}\sin i_* = 4.89 \pm 0.11$~kms$^{-1}$ and for run B $\lambda = 1.21\substack{+2.50 \\ -2.51}^{\circ}$ and $v_{\rm{eq}}\sin i_* = 4.95 \pm 0.11$~kms$^{-1}$. Overall, $v_{\rm{eq}}\sin i_*$ for both runs is consistent to better than 1$\sigma$ and the precision in $\lambda$ is the same. For the combined fit, $\lambda$ falls between the two values for both runs individually. The difference in $\lambda$ between the two runs is primarily driven by outlier observations at each limb with larger uncertainties. 

We run a SB model fit including an uncorrelated white noise term ($\sigma$) to check for any unaccounted for noise present in the data as a result of stellar surface variability, see Table \ref{mcmc_results}. From the model, $\sigma$ could potentially be picking up on p-modes oscillations which are present within the local RVs. As a result of this, we look to binning out these potential p-modes in Section \ref{clv}. 

\begin{figure}
    \centering
    \includegraphics[width = 0.49\textwidth]{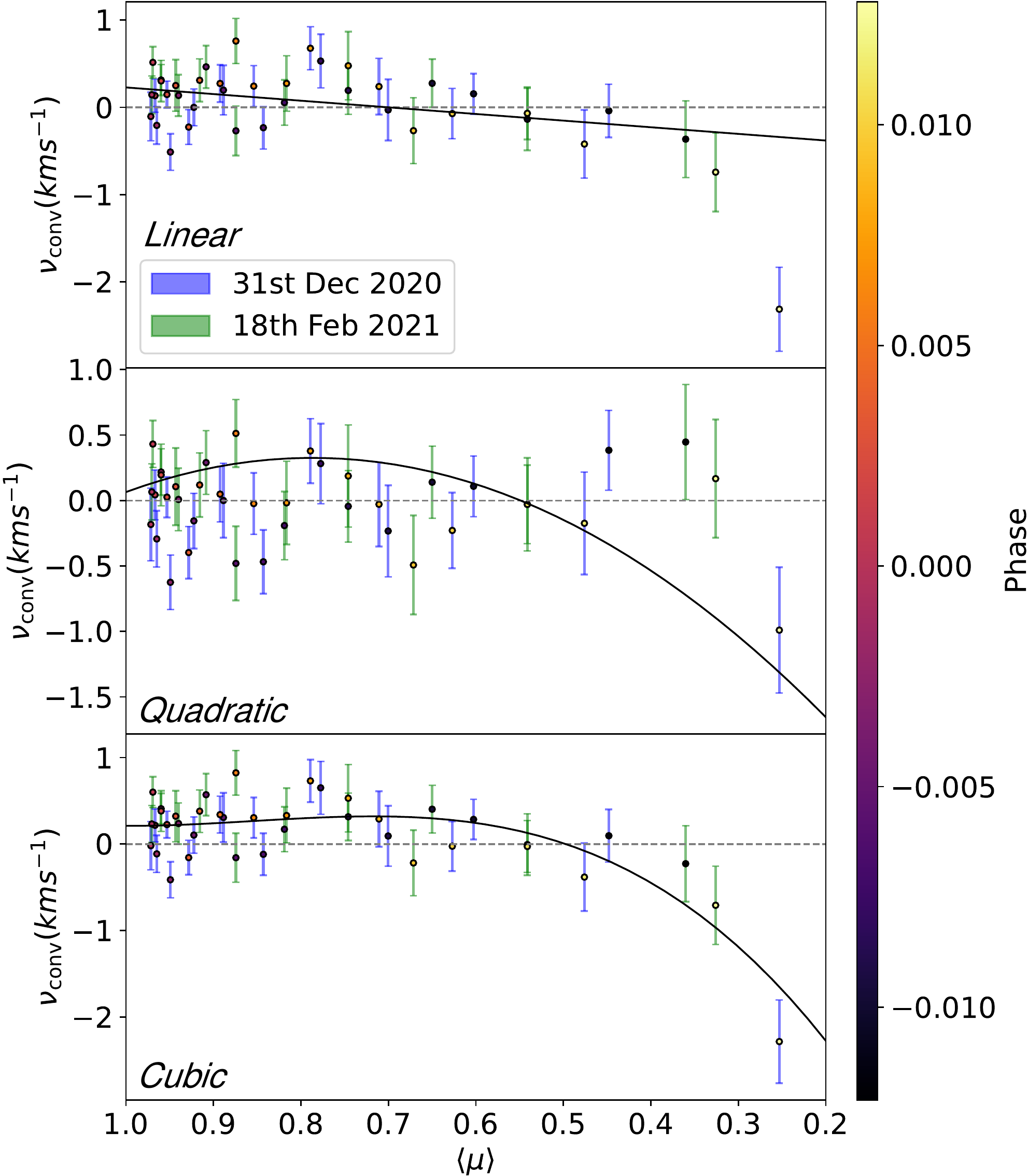}
    \caption{The net convective shifts determined by subtracting the solid body model fit (which changes slightly when adding in CLV) from the local RVs of the in-transit local CCFs, plotted as a function of stellar disk position behind the planet (brightness weighted $\langle\mu\rangle$). Model fits to the velocities are plotted as { \bf linear (top) and quadratic (middle) and cubic (bottom)}. The binned 10 minute data points are colour coded to indicate the phase with error bars colour coded according to the night of observation. Finally, the horizontal grey dashed lines at 0 are used to guide the eye.}
    \label{model:CLV}
\end{figure}

\subsection{Solid Body plus Centre-to-Limb Convective Variations Model}\label{clv}
We are also interested in how the net convective blueshift (CB) varies across the stellar disk. The net convective velocity shift caused by granules changes as a function of limb angle (i.e. from the centre to the limb of the star) due to line-of-sight (LOS) changes. Since WASP-166 is an F-type main sequence star (T$_{\rm{eff}}$ = 6050~K) we would also expect to observe the same phenomenon. To model the centre-to-limb convective variations (CLV) we fit the local RVs for CLV and SB at the same time. This was done by using $\lambda$ and $v_{\rm{eq}}\sin i_*$ for SB rotation and adding a linear, quadratic or cubic polynomial as a function of limb angle, see Table \ref{mcmc_results} and Figure \ref{model:CLV}. According to the BIC of the models, for the un-binned data, the SB plus quadratic CLV is the best fit to the data, resulting from a better fit to the data at disk centre. There is a marginal difference between the SB alone and SB plus linear CLV, with the BIC telling us the difference in fit is not sufficient enough to justify the extra free parameters. Additionally, there is a marginal difference between the SB plus quadratic and the SB plus cubic, where the chi-squared (\chisq) is slightly lower for the cubic fit but the BIC is higher indicating the extra free parameters are not justified. We would also like to point out that despite WASP-166~b being an aligned system (from the un-binned data), the local RVs at ingress and egress behave differently, see Figure \ref{local_RVs}. This is unexpected as we would expect to see the same behaviour as the planet transits the limb at ingress and egress. At present we do not have an explanation for this. 

\begin{figure*}
    \centering
    \subfloat{\includegraphics[width = 0.47\textwidth]{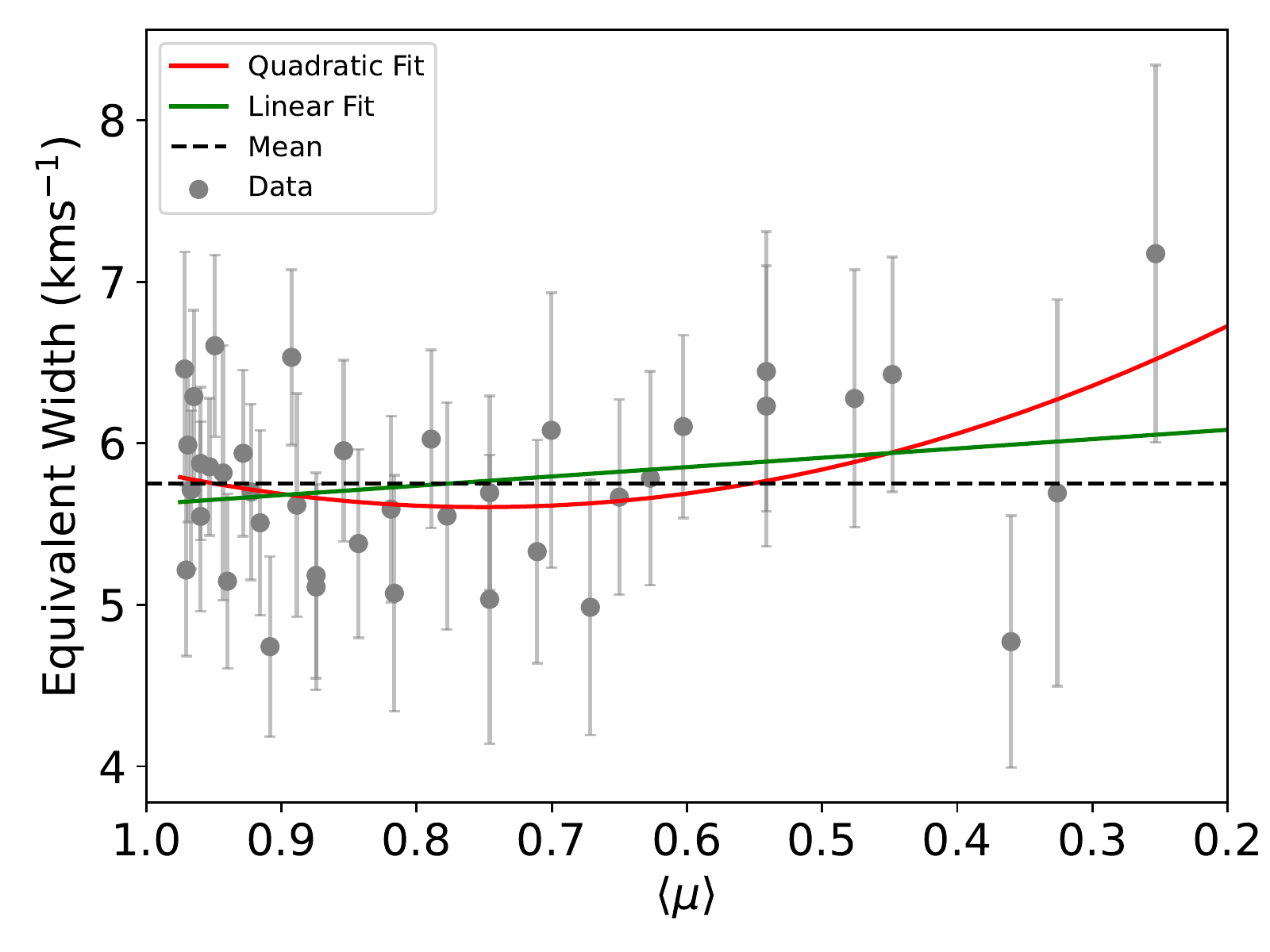}}
    \subfloat{\includegraphics[width = 0.47\textwidth]{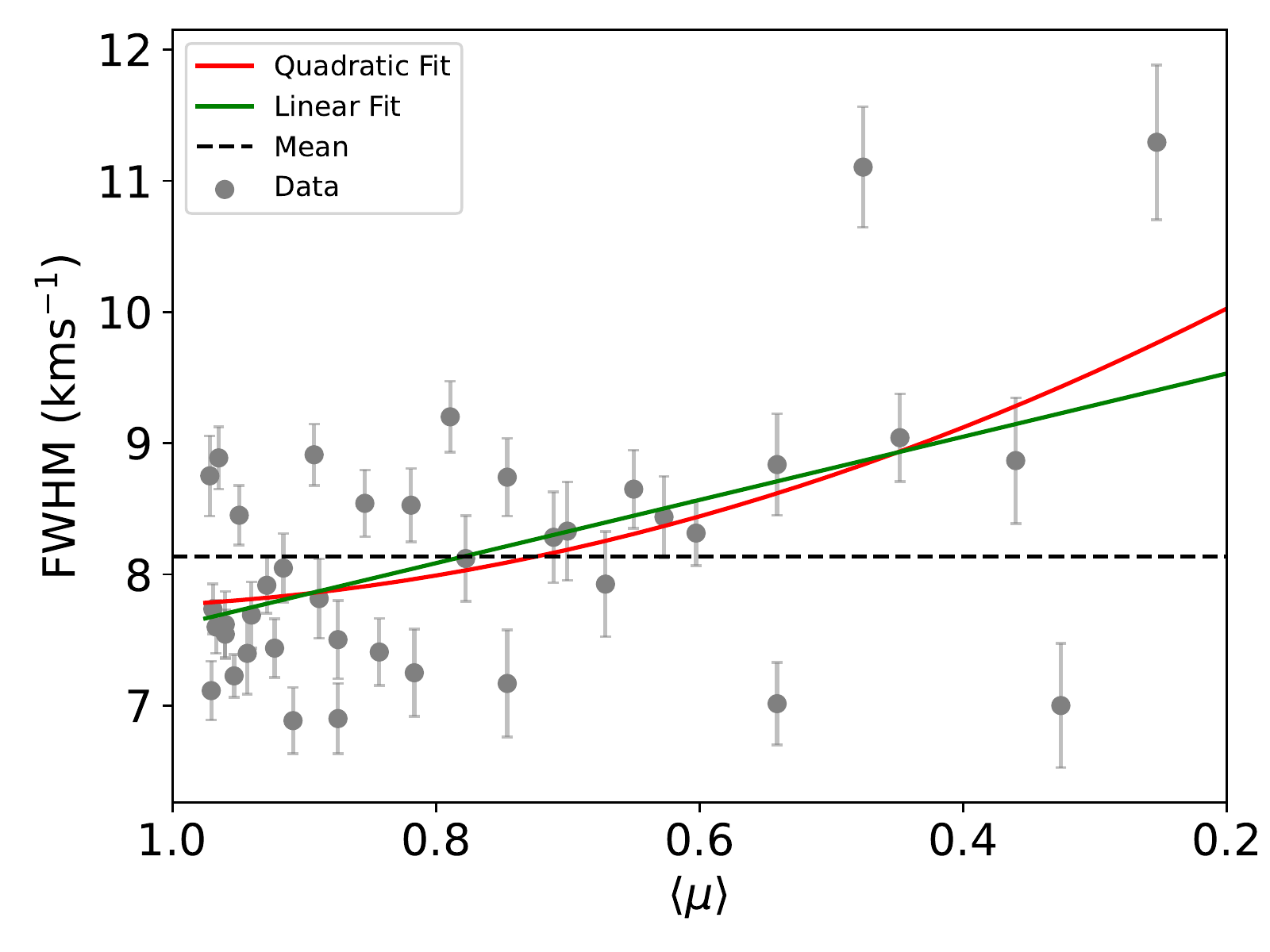}}
    \caption{The EW (left) and FWHM (right) of the \locccfs plotted as a function of stellar disk position behind the planet (brightness weighted $\langle\mu\rangle$). The grey data points represent the 10 min binned observations, the red line is a two degree polynomial fit to the data, the green line is a linear fit to the data and the dashed black line is the mean of the data.   }
    \label{ew_fwhm_ccf}
\end{figure*}

As mentioned previously we had potential p-mode oscillations in our local RVs. Radial velocity shifts caused by p-modes can be large enough to mask low mass, long and short period planetary signals. In our case, these p-modes are problematic as they introduce more noise and make it difficult to detect CLV. In \citet{chaplin2019filtering} they show how fine-tuning exposure times to certain stellar parameters can help to effectively average these oscillations. We used their {\tt Python} code {\tt ChaplinFilter} which outputs the ideal exposure times to use; (i) to suppress the total observed amplitude to a 0.1~ms$^{-1}$ level (29.8~mins) and (ii) corresponding to the Earth-analogue reflex amplitude of the star (21.4~mins). However, our RV precision for the ESPRESSO data is on a 1~ms$^{-1}$ level, therefore, to get to this we bin our RV errors to an exposure time of $\sim$10~mins. As a result of this, we binned the local RVs in time to increase our exposure time and increase our SNR, effectively averaging out the p-modes, by 10~mins. Additionally, we also binned the data to 20~mins which is between the {\tt ChaplinFilter} results and our ten minute bins as anymore binning would reduce our spatial resolution significantly. We refit all SB models to the 10 and 20~min binned local RVs, finding the uncorrelated white noise term ($\sigma$) drops to zero, supporting our initial suspicions of p-modes in the un-binned observations. 

For the 10 min binned data the best fit model to the data is the SB plus cubic CLV. This can be seen in Figure \ref{model:CLV}, where a cubic fit characterising the residual velocities (i.e. local RVs - SB model) is shown. The BIC for the quadratic CLV fit is the highest and, therefore, the worst model for the data. This can be seen clearly in Figure \ref{model:CLV} where the quadratic fit does not capture the shape of the data, however, the linear fit does capture this better in comparison. For the 20 min binned data, the SB alone is the best fit to the data with the lowest BIC value, however, the SB plus CLV cubic fit has a lower reduced chi-squared but the BIC is higher indicating the addition of free parameters is not justified. This is because this level of binning reduces the sampling of observations at the limb making it difficult to pick out any structure of the CLV. All the SB models for 10 and 20 min binned observations produce a consistent $v_{\rm{eq}}\sin i_*$ of approximately 4.80~kms$^{-1}$ and $\lambda$ within 2$\sigma$ of each other. When comparing all models from the un-binned observations, 10 min and 20 min binned observations, the reduced chi-squared ($\chi_{\nu}^{2}$, see Table \ref{mcmc_results}) indicates the SB plus cubic CLV model for the 10 min binned observations to be the best fit. All of the 1D distributions for each of the MCMC runs mentioned in this section can be found as supplementary material online. Therefore, the CLV is characterised by the cubic fit to have have a velocity of $\sim$ -1 to -2~kms$^{-1}$ at the limb (see Figure \ref{model:CLV}). In \citet{cegla2016modeling} they found for an aligned hot Jupiter with a four day orbit around a Sun-like star, if you neglect contributions of $\sim$5~kms$^{-1}$ in CLV there can be uncertainties in the project obliquity of $\sim$ 10 -- 20$^{\circ}$. For WASP-166, when comparing $\lambda$ between the SB model only and the SB plus cubic CLV of the binned 10 minute observations there is a difference of 9.6$^{\circ}$ which is in line with what \cite{cegla2016modeling} predicted. 

\subsection{Differential Rotation Models}
In the next scenario we apply the RRM model to the un-binned local RVs assuming differential rotation (DR) for the star. If DR is present, we can determine the latitudes on the star which are transited by the planet through disentangling $v_{\rm{eq}}\sin i_*$ and determine the 3D obliquity. Therefore, the MCMC parameters are $\alpha$, $\lambda$, $v_{\rm{eq}}$ and $i_*$ and the results can be found in Appendix \ref{corner:DR}. Unfortunately, we have bimodal distributions present in both $\alpha$ and $i_*$, informing us that there remains a degeneracy within them. This could be due to the spectroscopic transits not being precise enough to separate between if the star is pointing away or towards us. We also ran the MCMC fitting for DR fixing $i_*$ < 90$^{\circ}$ (away) and $i_*$ > 90$^{\circ}$ (towards) to get an estimate on $i_*$ and $\alpha$ finding these models had a higher BIC than the SB models. In \citet{hellier2019wasp} they estimate a stellar rotation period of 12.1~d from variations in the HARPS RVs and  combining the $v_{\rm{eq}}\sin i_*$ fitted to the RM effect with the fitted stellar radius (this was not done using photometry). From this we can disentangle $i_*$ from $v_{\rm{eq}}\sin i_*$ by using the rotation period and our $v_{\rm{eq}}\sin i_* =$ 4.89~kms$^{-1}$, yielding $i_*\sim$73.0$^{\circ}$. We can then use this to predict the stellar latitudes transited by the planet, using P$_{\rm{rot}}$ and $v_{\rm{eq}}\sin i_*$, which gives a $\sim$9$^{\circ}$ change in latitude. 

In \citet{roguetkern2021drclv} they investigate the optimal parameter space to use the RRM technique to detect DR and CLV on a HD~189733-like system (i.e. a hot Jupiter in a circular orbit around a K-dwarf). To do this they use simulations to explore all possible ranges of $\lambda$, $i_*$ and impact factor, $b$, producing maps of optimal regions. We place WASP-166 with $\lambda = -4.49^{\circ}$ (from SB model, un-binned), $i_* = 88.0^{\circ}$ \citep[from][estimation]{hellier2019wasp} and $b = 0.213$ (from Table \ref{planetary_properties}) on the heat maps in their Figure 3. According to their maps we find the chance of detecting DR on WASP-166 is slim given these parameters with a change in BIC of $<$2. We ran the DR plus CLV linear, quadratic and cubic models on the 10 minute binned observations to see if accounting for CLV made a difference when trying to pull out DR, see Table \ref{mcmc_results}. In all of the models there was a bimodal distribution present in $i_*$, suggesting we cannot discern if the star is pointing away or towards us. Furthermore, the BIC between the SB plus cubic CLV and DR plus cubic CLV has a difference of $\sim$1, which isn't a significant enough difference for the DR model to be preferred. Additionally, the differences in \chisq and BIC are minimal between the away and towards models meaning we cannot discern between them, see Section \ref{dr_analysis} for full details. Therefore, the best model to fit the data is the SB plus cubic CLV.

\subsection{Variations within the CCF profiles}
In section \ref{clv} we modelled the local RVs from fitting the local CCFs. However, in this section we take a closer inspection of the CCFs themselves and how their shape changes as a function of limb angle. We computed an analysis into the equivalent width (EW) and FWHM (as they represent a measure of changes in the width and height of the profile) of our CCF profiles, using the 10 min binned observations (due to the increase in SNR) from the previous section. In Figure \ref{ew_fwhm_ccf}, we show the EW and FWHM plots for the 10 min binned data calculated from the Gaussian fitting method explained in Section \ref{get_vel}. For EW we observe a trend which increases towards the limb and the same for FWHM. We fit both a linear and quadratic relationship to each and derive the R$^2$ and p-value as a measure of the goodness of fit. The R$^2$ measures the degree to which the data is explained by the model, where a higher value indicates a better fit. The p-value then indicates if there is enough evidence that the model explains the data better than a null model (i.e. if the p-value is very small, we can reject the null hypothesis). For the EW, R$^2$ = 0.04 and p-value = 0.18 for a linear fit as a function of $\mu$ and R$^2$ = 0.12 and p-value = 0.09 for a quadratic fit, where a quadratic is the better fit to the data. Similarly, for the FWHM, R$^2$ = 0.24 and p-value = 0.001 for a linear fit and R$^2$ = 0.25 and p-value = 0.004 for a quadratic fit. In this case, the p-value is lower for the linear model despite the R$^2$ being lower in comparison to the quadratic model. Both models are a good fit to the data, however, we will select the model with the lower p-value which has fewer free parameters as we do not need to over complicate the model. Therefore, we argue that the linear model is the best fit to the FWHM data. In \citet{beeck2013three} they use 3D HD simulations to model the convection in a range of cool main sequence stars, then use these models to synthesise various Fe~I line profiles. They found similar trends in EW and FWHM for their models of F- (T$_{\rm{eff}} \sim$~6850~K) and G-type (T$_{\rm{eff}} \sim$~5800~K) stars. They attribute the increase in EW towards the limb to be a result of the Fe~I lines being stronger due to the increasing temperature of the lower photosphere with respect to optical depth (at the limb our line-of-sight means we see more of the granular walls). For FWHM, an increase towards the limb was observed for all stars in their models due to the increasing component of horizontal velocity flows compared to line-of-sight velocity causing a stronger Doppler broadening. For WASP-166, we observe the same behaviour as \citet{beeck2013three}. They synthesise and analyse the Fe~I line at 617.3~nm only, however, we are looking at the CCF which is an average of all absorption lines within the CCF template mask including a large number of Fe lines. In a more recent study, \citet{dravins2018spatially} investigate the centre to limb changes of line widths in HD~189733, a cool K-dwarf, finding the lack of any corresponding observed signature/trend. However, they go on to model the changes in line width for F, G and K-type stars as a function of limb angle (see their Fig 11) finding a change in the region of $\sim$3~kms$^{-1}$. In another study \citep{dravins2017spatially}, they compute a similar analysis on the hotter G-type star HD-209458 ($T_{\rm{eff}} = $ 6071~K) finding a change in velocity of between 1 -- 2~kms$^{-1}$ in line width towards the limb (see their Figure 17), which is comparable to our measurements in Figure \ref{ew_fwhm_ccf}. Overall, our finding using observations solidifies that of \cite{beeck2013three}, \citet{dravins2018spatially} and \citet{dravins2017spatially} who used simulations.  

\begin{figure}
    \centering
    \includegraphics[width = 0.47\textwidth]{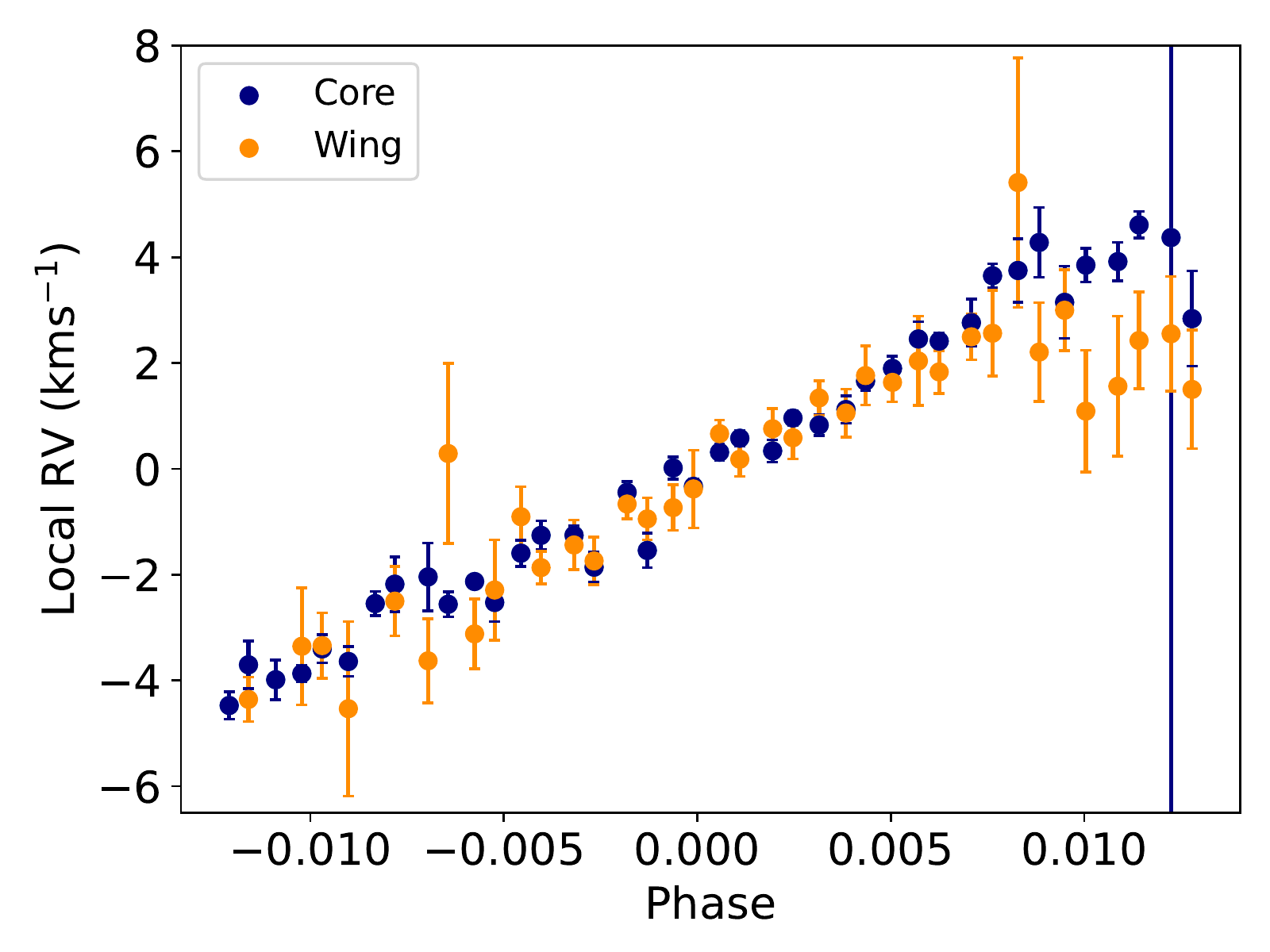}
    \caption{The local RVs for the Gaussian fits to the CCF core (blue) and wing (orange) for the ten minute binned observations.  }
    \label{RVs_wing_core}
\end{figure}

In addition to looking at the EW and FWHM we can also isolate different parts of the CCF to fit individually. For this analysis we focus on fitting the CCF core and CCF wings as each originate from a different depth in the photosphere. We expect asymmetries within the CCFs as we move from disk centre to the limb caused by the net velocities of the granules on the surface of the star. We define the CCF wing to be 50\% of the line depth including the continuum and the CCF core to be the remaining 50\% of the line. Due to the sampling of the CCF this was needed in order to have enough data points for a sufficient Gaussian profile to be fitted. The Gaussian fitting was done using the same method described in Section \ref{get_vel} with no constraint placed on the continuum of the line core fit. In Figure \ref{RVs_wing_core}, we show the local RVs for the CCF core and CCF wing of the binned 10 minute observations where a split can be seen towards the limb of the stars between the two at the end of the transit. This shows there are asymmetries within the CCF with the wings appearing more redshifted towards the limb at the end of the transit than the core. We also fitted the local RVs of the wing and core using the MCMC method described in Section \ref{get_vel} and the results are in Table \ref{mcmc_results}. The asymmetries in the line produced a rather high $\lambda$ for the wing, whereas for the core $\lambda$ remained consistent with an aligned system. We investigated the residual RVs (subtracting the local RVs from SB model) however, nothing of significance was found. 

\section{Conclusions}
We have conducted an in depth study into the  stellar surface variability of WASP-166 while also investigating the impacts this has on the projected obliquity. To do this, we utilised photometric transits of WASP-166~b, two observed with NGTS (simultaneous to ESPRESSO) and a further six observed with {\tess}, to fit for the transit planetary system properties (see Table \ref{planetary_properties}). We then computed the Reloaded Rossiter McLaughlin (RRM) analysis using two full ESPRESSO transits to determine the local velocity of the star behind the planet, testing these RVs for various models to include solid body (SB) rotation, centre-to-limb convective variations (CLV), and differential rotation (DR), see Table \ref{mcmc_results}. From a SB model fit of the un-binned ESPRESSO observations we found WASP-166 has a projected obliquity of $\lambda = -4.49^{+1.74}_{-1.73}$$^{\circ}$ and $v_{\rm{eq}}\sin i_* = 4.89 \pm 0.08$~kms$^{-1}$ which is consistent with WASP-166~b being an aligned system. However, there was evidence of p-modes within our dataset so, we binned the data by 10 minutes to achieve an RV precision of $\sim$1~ms$^{-1}$ (this is the expected precision from the exposure time calculator for ESPRESSO). This method effectively averages out the p-modes which can introduce more noise and make it difficult to detect CLV. Fitting the 10 minute binned observations with a SB RRM model yielded $\lambda = -5.93^{+1.99}_{-2.01}$$^{\circ}$ and $v_{\rm{eq}}\sin i_* = 4.77 \pm 0.09$~kms$^{-1}$ which remains consistent with our un-binned analysis and WASP-166~b being an aligned system. This is an update on previous studies of WASP-166~b \citep{hellier2019wasp, vedad2021wasps} using HARPS observations where we are in agreement with their values, however, in this study the ESPRESSO observations allow us to be more precise. 

In addition to fitting for SB we also looked at models with CLV. The net convective velocity from the centre of the star out to the limb changes due to the convective cells being viewed at different angles from changes in LOS. Therefore, we alter the RRM model to account for this by fitting a linear, quadratic and cubic relation as a function of limb angle. In the 10 minute binned dataset the SB plus cubic CLV model was the best fit to the data with an improvement in the BIC of twenty (see Table \ref{mcmc_results} and Figure \ref{model:CLV} for the plotted trends). The $\Delta$BIC (difference between a particular model and the best model with the lowest BIC) can be used as an argument against the other model. A $\Delta$BIC between 2 and 6 means there is a good argument in favour of the best model and between 6 and 10 the evidence for the best model and against the weaker model is strong. We used an Anderson Darling test to check if the residuals for the un-binned observations in Figure \ref{local_RVs} were normally distributed and found they were not indicating there is remaining correlated red noise as a result of magneto convection (i.e. granulation/super granulation). Therefore, we would like to caution the reader on the interpolation of the BIC of the models. For the 20 minute binned observations, the SB alone is the best fit to the data due to the level of binning reducing the sampling of observations at the limb, making it difficult to pick out any structure of the CLV. Overall, ignoring the velocity contributions induced by granulation can skew the measurements of the projected obliquity and, therefore, impact out understanding of planet formation and evolution of the system. This can be seen in our results within this paper where the projected obliquity changes by 9.6$^{\circ}$ for 10 minute observations when CLV are accounted for. However, in the case of WASP-166 this change still means the system is aligned and as a result will not change our view of planet formation and evolution. For the un-binned observations $\lambda = -4.49^{+1.74}_{-1.73}$ and in 10 minute binned observations $\lambda = -15.52^{+2.85}_{-2.76}$, $v_{\rm{eq}}\sin i_*$ remains consistent within 2$\sigma$ of the SB model results. This is the first time where a tentative CLV has been detected on WASP-166 using the RRM technique. 

We attempted to fit a DR model to the un-binned data, however, due to the number of stellar latitudes transited by WASP-166~b and our RV precision we were unable to constrain any DR for this data set. Additionally, we were unable to disentangle $i_*$ from $v_{\rm{eq}}\sin i_*$ as we could not determine a rotation period for WASP-166 from the available photometric data. From our predictions of the local RVs and keeping $\lambda = -4.49^{\circ}$, to pull out DR WASP-166 would need to posses an $i_*$ between 110$^{\circ}$ and 160$^{\circ}$ or $<$ 40$^{\circ}$ \citep[from][]{roguetkern2021drclv}, where more stellar latitudes would be crossed by the planet and the local RV variation induced by the DR would be greater than the uncertainties. However, we determine a CLV between $\sim$ -1 -- -2~kms$^{-1}$, therefore, by including this in our modelling we were able to fit and pull out a potential DR shear of $\alpha$ $\sim$ -0.5 (anti-solar) and $i_*$ either 42.03$^{+9.13}_{-9.60}$$^{\circ}$ or 133.64$^{+8.42}_{-7.98}$$^{\circ}$ (i.e. either pointing away from or towards us). This further strengthens the need to consider and fit for CLV as it can also impact the detection and characterisation of DR as it is may be masked out by the CLV.

Finally, we also inspected how the shape of the CCF changes as a function of limb angle. We found that the FWHM and equivalent width (EW) of the local CCFs for the 10 minute binned data show trends which increase towards the limb of the star by $\sim$2~kms$^{-1}$. We modelled this with a linear and quadratic fit, using the R$^2$ value to determine the best fit was a quadratic for EW and linear for FWHM. This analysis matches the findings of both \citet{beeck2013three} and \cite{dravins2017spatially, dravins2018spatially} who found similar results in simulated line profiles from state-of-the-art 3D HD simulations. Therefore, these types of observations allow us to further validate these simulations for main sequence stars other than the Sun. Finally, we looked at the asymmetry in the local CCFs along the transit chord by fitting for the core and wing separately, finding the wing was more redshifted than the core. This has been seen for the Sun using HD simulations of the Fe~I line in \citet{cegla2018stellar} (see Figure 11) where there is a C-shape within the line bisector which becomes more prominent towards the solar disk. Near the limb we lose the downward bend in the wing which is believed to be the contribution from the down flowing intergranular lanes.

Overall, for WASP-166 we have refined the system parameters and investigated the stellar surface variability, comparing this to and validating 3D HD simulations. We find a centre-to-limb convective variation detection which has an impact on deriving the projected obliquity. Furthermore, by including the centre-to-limb contribution and modelling it together with differential rotation were were able to put limits on the differential rotational shear and stellar inclination.  

\section*{Acknowledgements}
This work is based on observations made with ESO Telescopes at the La Silla Paranal Observatory under the programme ID 106.21EM. We also include data collected by the {\tess} mission, where funding for the {\tess} mission is provided by the NASA Explorer Program. LD, HMC and ML acknowledge funding from a UKRI Future Leader Fellowship, grant number MR/S035214/1. This work has been carried out in the frame of the National Centre for Competence in Research PlanetS supported by the Swiss National Science Foundation (SNSF). This project has received funding from the European Research Council (ERC) under the European Union's Horizon 2020 research and innovation programme (project {\sc Spice Dune}, grant agreement No. 947634). RA is a Trottier Postdoctoral Fellow and acknowledges support from the Trottier Family Foundation. This work was supported in part through a grant from FRQNT. VK acknowledges support from NSF award AST2009501. JSJ gratefully acknowledges support by FONDECYT grant 1201371 and from the ANID BASAL projects ACE210002 and FB210003.

\section*{Data Availability}
The {\tess} data are available from the NASA MAST portal and the ESO ESPRESSO data are public from the ESO data archive. NGTS data is available for the two simultaneous transits as supplementary material online with this paper. The third NGTS transit is available to download online from \citet{bryant20multicam}.



\bibliographystyle{mnras}
\bibliography{wasp-166} 

\appendix

\section{NGTS Photometric Data}
\begin{table}
	\centering
	\caption{Example table of the NGTS photometry of WASP-166 obtained for this study on the nights 2020 December 31 and 2021 February 18. The full table is available online.}
	\label{tab:phot}
\begin{tabular}{cccc}
    BJD (TDB) &             Flux &  Flux Error & Cam\\
  (-2,450,000) &   &   &  \\ \hline
       9215.63952437 & 1.000649 & 0.014741 & 1\\
       9215.63968642 & 1.001084 & 0.014700 & 1\\
       9215.63983689 & 0.998373 & 0.014646 & 1\\
       9215.63998737 & 1.010102 & 0.014726 & 1\\
       9215.64013784 & 0.987914 & 0.014447 & 1\\
       9215.64027674 & 1.019617 & 0.014732 & 1\\

\hline
	\end{tabular}
\end{table}

\section{Full TESS Lightcurves}
\begin{figure*}
    \centering
    \includegraphics[width=0.97\textwidth]{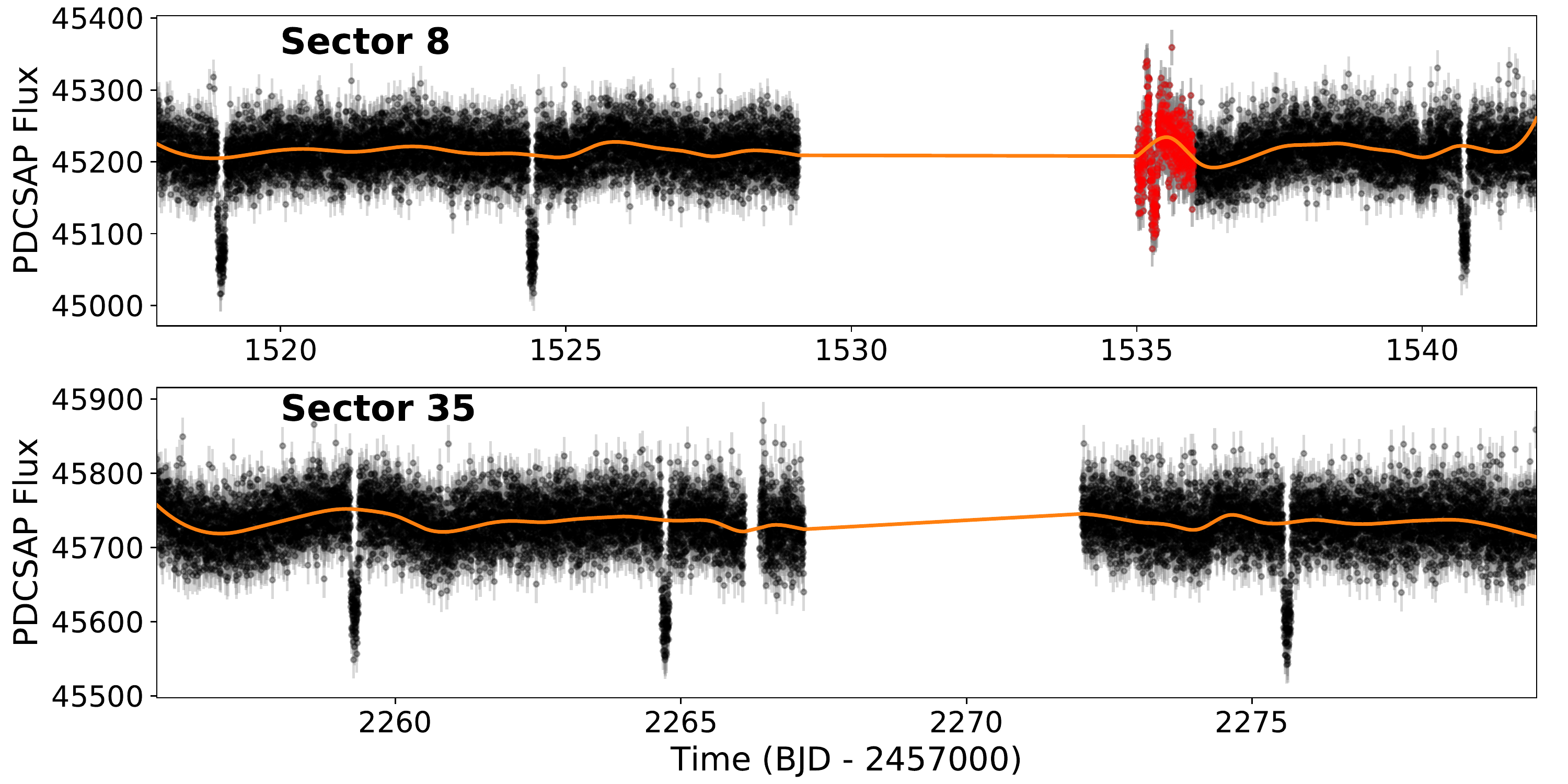}
    \caption{Full PDCSAP flux time series for the TESS photometry of WASP-166 from Sectors 8 and 35. The orange line shows the spline fit used to flatten the data (see Section~\ref{sec:tessphot}) and the red points highlight the region of data at the beginning of the second Sector 8 orbit excluded from our analysis (see Section~\ref{sec:trans_analysis}).}
    \label{fig:tess_LC}
\end{figure*}

\section{Differential Rotation plus Centre-to-Limb Convective Variations Model}
\label{dr_analysis}
In Section \ref{clv} when modelling the SB rotation scenario we find the CLV could potentially be on the order of between $\sim$ -1 -- -2~kms$^{-1}$. This is such a large additional velocity contribution it could potentially mask the DR velocity signature. For example, if WASP-166~b crossed all latitudes of WASP-166 and the star possessed a differential rotational shear ($\alpha$) of 0.5, then the change in velocity from pole to pole would be $\sim$2.5~kms$^{-1}$. This is very close to the value we estimate the CLV to be, therefore, it may be possible to pull out DR if we account for CLV. As a result, we ran the DR plus CLV linear, quadratic and cubic models on the 10 minute binned observations, see Table \ref{mcmc_results}. In all of these models there was a bimodal distribution present in $i_*$, suggesting we cannot discern if the star is pointing away or towards us (see the supplementary online material for the MCMC distributions). Overall, we find the DR plus CLV cubic model fits the data nearly as well as our SB plus CLV cubic model with a $\Delta$BIC $\sim$ 1 larger. The derived differential rotation shear is between -0.48 and -0.58 which corresponds to anti-solar differential rotational shear (where the poles rotate faster than the equator). For cool main sequence stars, such as WASP-166, anti-solar DR has been linked to slow rotators with a high Rossby number. In \citet{barnes2005dr} they conclude DR decreases with effective temperature and rotation. In a more recent study, \citet{kHovari2017rotation} suggested a DR rotation period relationship from Doppler imaging results. Despite this, \citet{reinhold2013rotation} observed a rather weak variation of the observed differential rotation on the rotation periods of cool stars (between T$_{\rm{eff}}$ of 3500 to 6000~K), with stars above 6000~K showing no systematic trend and being randomly distributed within the temperature regime (note: WASP-166 has T$_{\rm{eff}}$ = 6050~K and is F9 spectral type). Additionally, hotter stars show stronger differential rotation which peaks at F-type stars, consistent with our findings here for WASP-166. Finally, \citet{benomar2018asteroseismic} use asteroseismology to detect DR in a sample of solar-type stars and found 10 out of 40 in their sample were  anti-solar and, overall, a weak anti-correlation to rapid rotation. Therefore, our finding of $\alpha \sim$ -0.5, is quite plausible meaning WASP-166 might have a high anti-solar DR shear.

\section{MCMC Posterior Probability Distributions }

\begin{figure*}
    \centering
    \includegraphics[width = 0.70\textwidth]{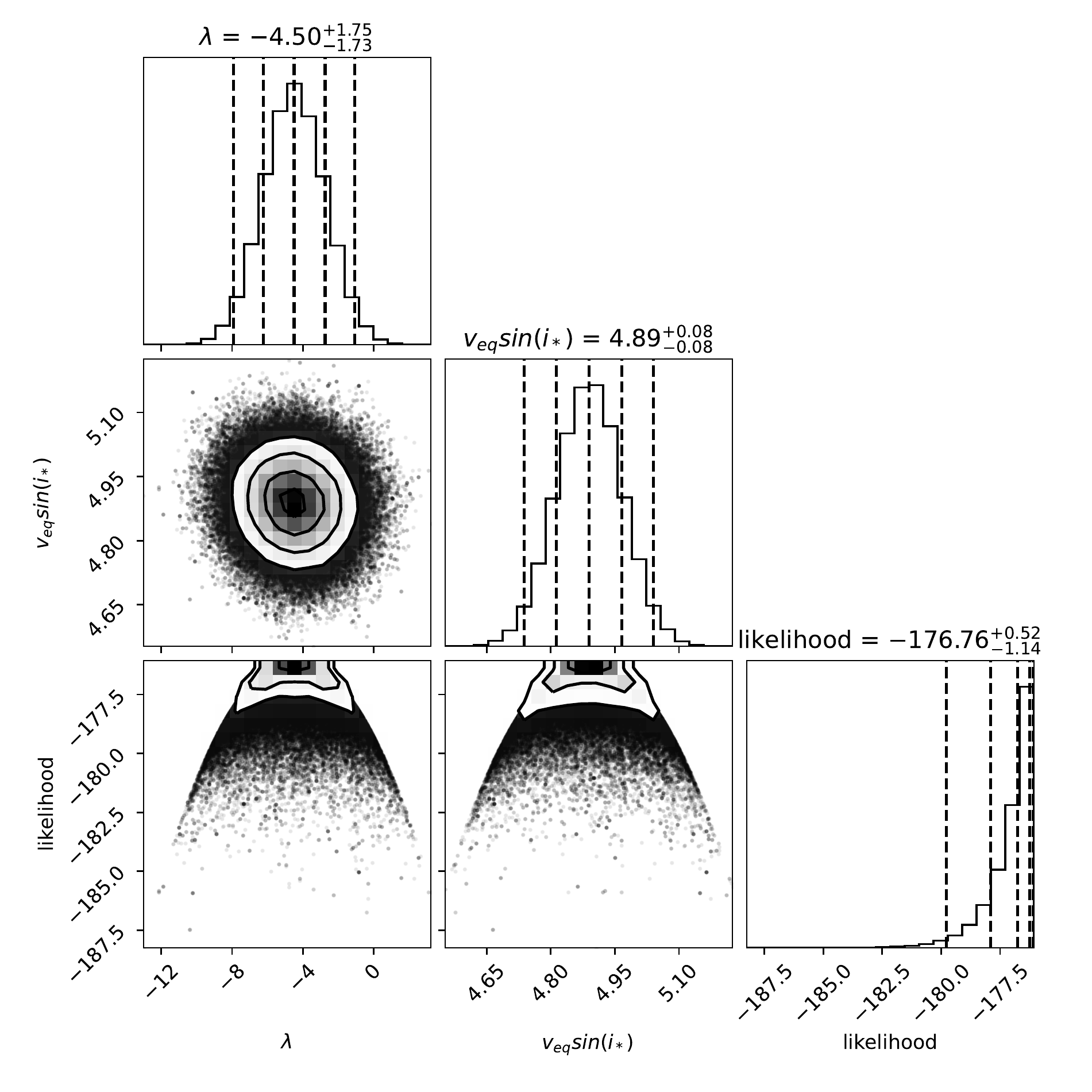}
    \caption{The corner plot of the solid body MCMC showing the one and two dimensional projections of the posterior probability distribution for the parameters. The vertical dashed lines represent the 16th, 50th and 84th percentiles of each parameter.}
    \label{corner:SB}
\end{figure*}

\begin{figure*}
    \centering
    \includegraphics[width = 0.90\textwidth]{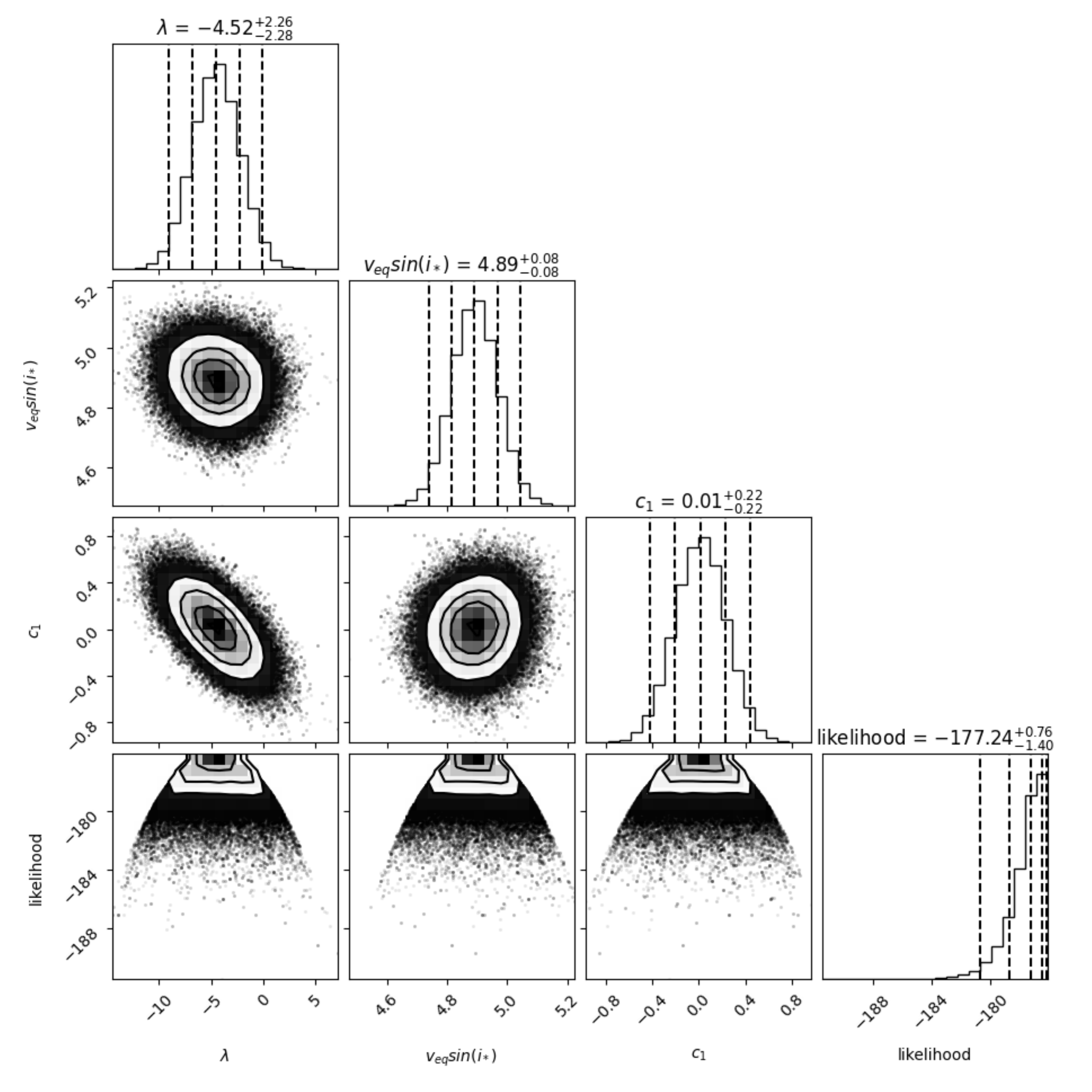}
    \caption{The corner plot of the solid body and linear centre-to-limb convective variations MCMC showing the one and two dimensional projections of the posterior probability distribution for the parameters. The vertical dashed lines represent the 16th, 50th and 84th percentiles of each parameter.}
    \label{corner:SB_clv1}
\end{figure*}

\begin{figure*}
    \centering
    \includegraphics[width = 0.97\textwidth]{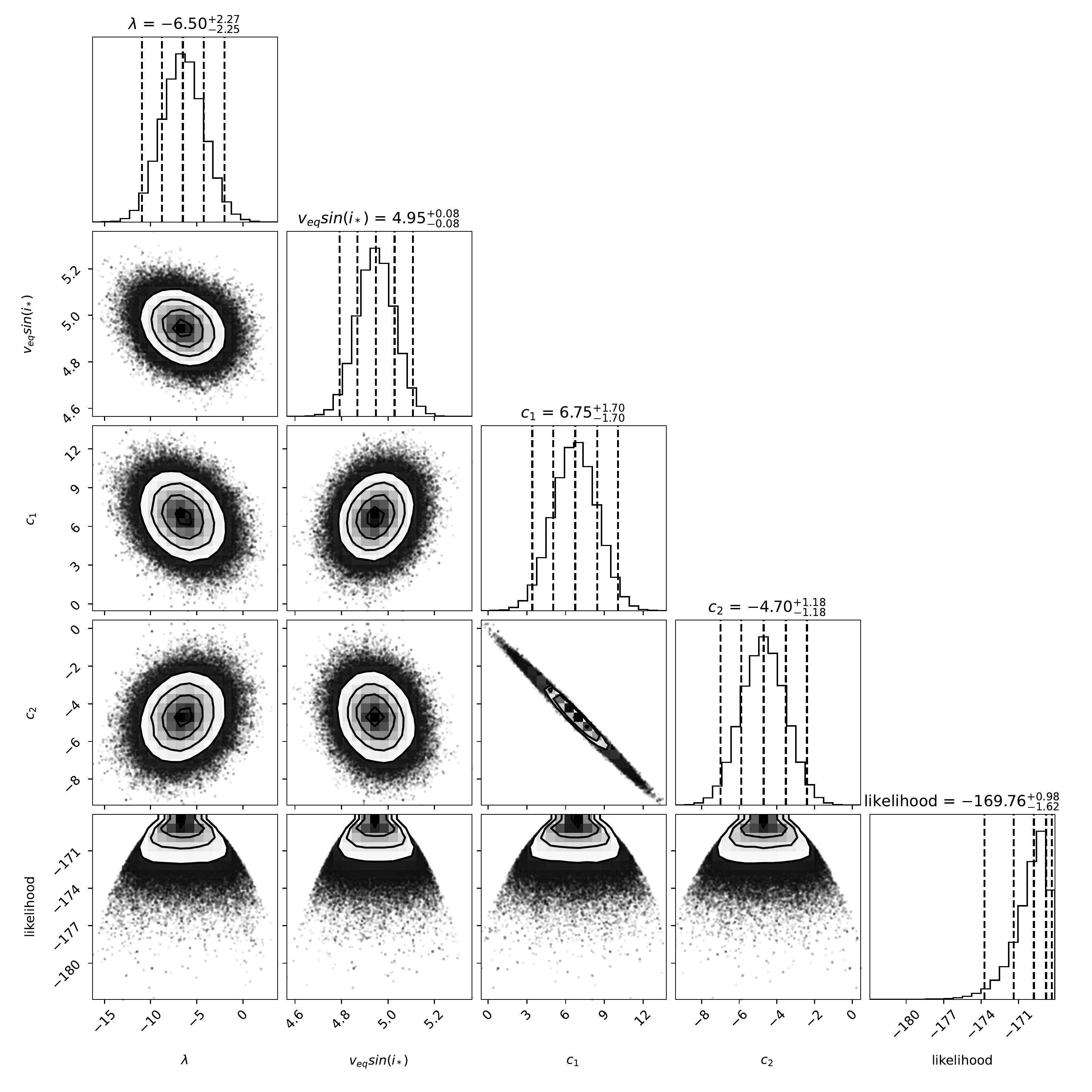}
    \caption{The corner plot of the solid body and quadratic centre-to-limb convective variations MCMC showing the one and two dimensional projections of the posterior probability distribution for the parameters. The vertical dashed lines represent the 16th, 50th and 84th percentiles of each parameter.}
    \label{corner:SB_clv2}
\end{figure*}

\begin{figure*}
    \centering
    \includegraphics[width = 0.97\textwidth]{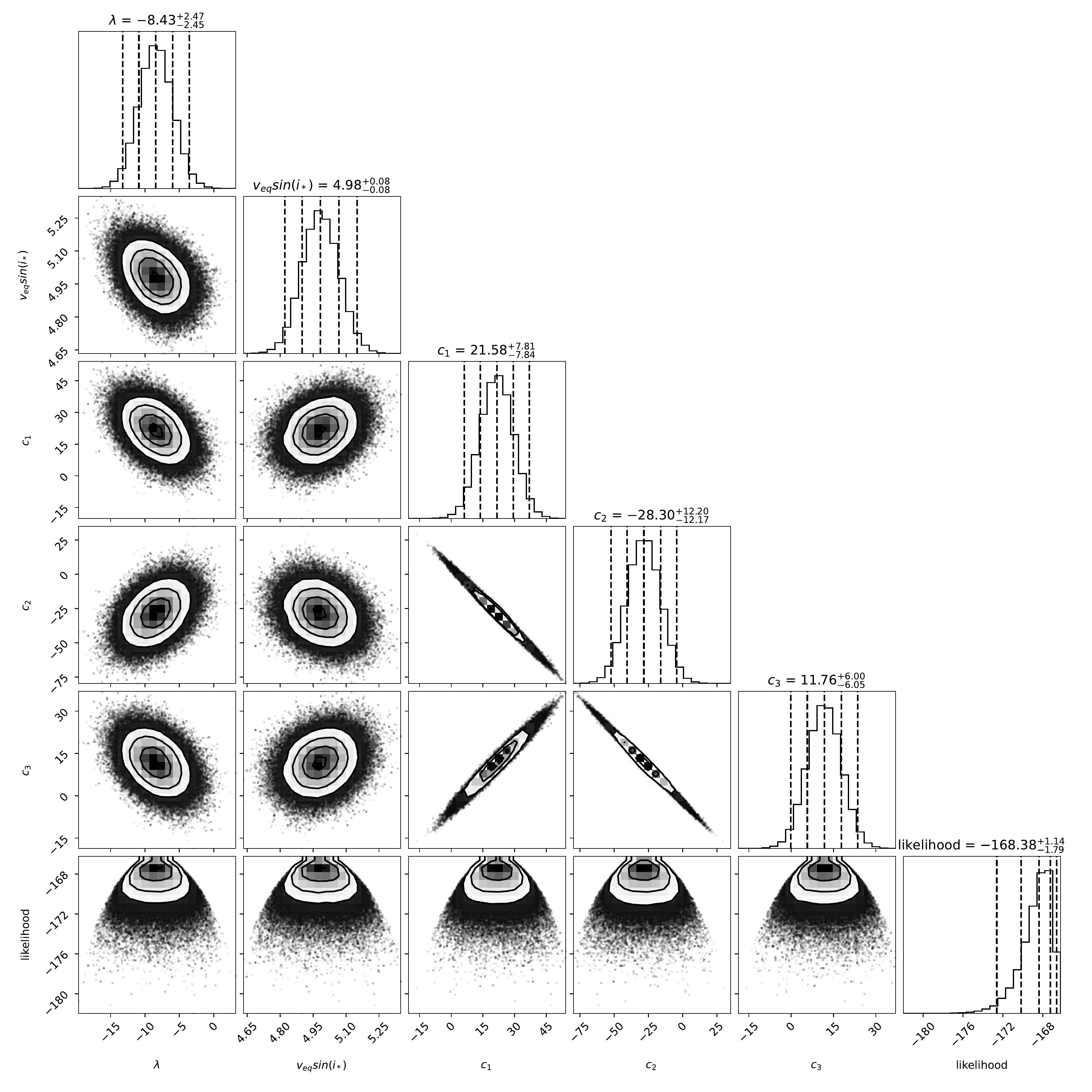}
    \caption{The corner plot of the solid body and cubic centre-to-limb convective variations MCMC showing the one and two dimensional projections of the posterior probability distribution for the parameters. The vertical dashed lines represent the 16th, 50th and 84th percentiles of each parameter.}
    \label{corner:SB_clv3}
\end{figure*}

\begin{figure*}
    \centering
    \includegraphics[width = 0.97\textwidth]{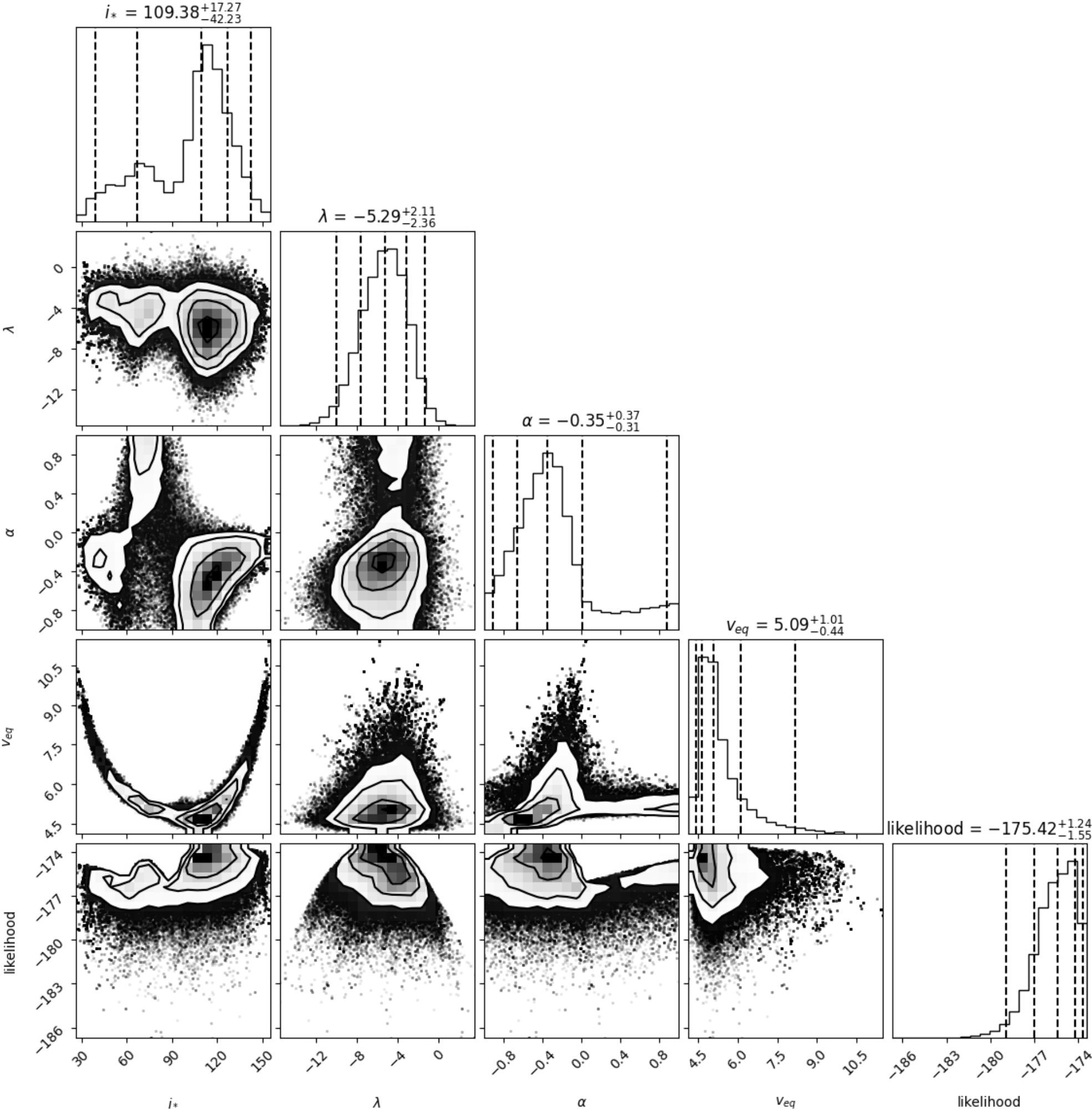}
    \caption{The corner plot of the differential rotation MCMC showing the one and two dimensional projections of the posterior probability distribution for the parameters. The vertical dashed lines represent the 16th, 50th and 84th percentiles of each parameter.}
    \label{corner:DR}
\end{figure*}

\bsp	
\label{lastpage}
\end{document}